\documentclass[iop,apj,twocolappendix]{emulateapj}
\usepackage[varg]{txfonts}
\usepackage{bm}
 \usepackage{color}

\newcommand{\beq}{\begin{equation}}
\newcommand{\eeq}{\end{equation}}
\newcommand{\beqn}{\begin{eqnarray}}
\newcommand{\eeqn}{\end{eqnarray}}
\newcommand{\pd}{\partial}
\newcommand{\AU}{{\rm AU}} 

\newcommand{\kB}{k_{\rm B}}

\newcommand{\ovl}[1]{ {\overline{#1}} }

\newcommand{\bracket}[1]{\langle #1 \rangle}

\newcommand{\eqref}[1]{(\ref{#1})}

\newcommand{\dfrac}[2]{ {\displaystyle\frac{#1}{#2}} }
\newcommand{\pfrac}[2]{ \biggl(\dfrac{#1}{#2}\biggr) }

\newcommand{\eps}{\epsilon}
\renewcommand{\leq}{\leqslant}
\renewcommand{\geq}{\geqslant}
\newcommand{\calN}{{\cal N}}

\shorttitle{RAPID COAGULATION OF POROUS DUST AGGREGATES OUTSIDE THE SNOW LINE}
\shortauthors{OKUZUMI ET AL.}
\slugcomment{ApJ, in press}

\begin{document}
\title{Rapid Coagulation of Porous Dust Aggregates Outside the Snow Line: A Pathway to
Successful Icy Planetesimal Formation}
\author{Satoshi Okuzumi\altaffilmark{1}, Hidekazu Tanaka\altaffilmark{2}, 
Hiroshi Kobayashi\altaffilmark{1}, and Koji Wada\altaffilmark{3}
}
\altaffiltext{1}{Department of Physics, Nagoya University, Nagoya, Aichi 464-8602, Japan; \email{okuzumi@nagoya-u.jp}}
\altaffiltext{2}{Institute of Low Temperature Science, Hokkaido University, Sapporo 060-0819, Japan}
\altaffiltext{3}{Planetary Exploration Research Center, Chiba Institute of Technology, Narashino, Chiba 275-0016, Japan}

\begin{abstract}
Rapid orbital drift of macroscopic dust particles is one of the major obstacles against planetesimal formation in protoplanetary disks. We reexamine this problem by considering porosity evolution of dust aggregates. We apply a porosity model based on recent $N$-body simulations of aggregate collisions, which allows us to study the porosity change upon collision for a wide range of impact energies. As a first step, we neglect collisional fragmentation and instead focus on dust evolution outside the snow line, where the fragmentation has been suggested to be less significant than inside the snow line because of a high sticking efficiency of icy particles. We show that dust particles can evolve into highly porous aggregates (with internal densities of much less than $0.1~{\rm g~cm^{-3}}$) even if collisional compression is taken into account. We also show that the high porosity triggers significant acceleration in collisional growth. This acceleration is a natural consequence of particles' aerodynamical property at low Knudsen numbers, i.e., at particle radii larger than the mean free path of the gas molecules. Thanks to this rapid growth, the highly porous aggregates are found to overcome the radial drift barrier at orbital radii less than 10 AU (assuming the minimum-mass solar nebula model). This suggests that, if collisional fragmentation is truly insignificant, formation of icy planetesimals is possible via direct collisional growth of submicron-sized icy particles. 
\end{abstract}
\maketitle

\section{Introduction}\label{sec:intro}
Growth of dust particles is a key process in protoplanetary disks.
Current theories of planet formation assume kilometer-sized solid bodies called ``planetesimals''
 to form from dust contained in protoplanetary disks.
Being the dominant component of disk opacity, dust also affects the temperature and observational appearance of the disks.
Furthermore, dust particles are known to efficiently capture ionized gas particles in the gas disk,
thereby controlling magnetohydrodynamical behaviors of it \citep{SMUN00}.

Theoretically, however, it is poorly understood how the dust particles evolve into planetesimals.
One of the most serious obstacles is the radial inward drift of macroscopic aggregates due to 
the gas drag \citep{W72,AHN76,W77}.
Because of the gas pressure support in addition to the centrifugal force, 
protoplanetary disks tend to rotate at sub-Keplerian velocities.
By contrast, dust particles are free from the pressure support, and hence
they tend to rotate faster than the gas disk.
The resulting head wind acting on the dust particles extracts their angular momentum
and thus causes their drift motion toward the central star.
In order to go beyond this ``radial drift barrier,'' dust particles must 
decouple from the gas drag (i.e., grow large) faster than they drift inward. 
However, previous work by \citet{BDH08} showed that dust particles finally fall onto the central star
unless the initial dust-to-gas mass ratio is considerably higher than
the canonical interstellar value.

Several mechanisms have been raised so far regarding how 
dust particles overcome the radial drift barrier.
A classical idea is that dust particles ``jump'' across the barrier by forming a gravitationally 
unstable thin dust layer at the midplane and directly collapsing into planetesimal-size objects
\citep{S69,GW73,HNN85}.
However, this classical scenario has been challenged 
by the fact that the dust layers are easily stirred up by disk turbulence \citep{WC93,TCS10}.
Moreover, the dust sublayer is known to induce the Kelvin-Helmholtz instability, 
which prevents further sedimentation of dust even without disk turbulence 
unless the dust-to-gas surface density ratio is considerably high \citep{S98}.
Recently, a two-fluid instability of dust and gas has been discovered \citep{YG05}, 
which can lead to fast formation of gravitationally bound dust clumps \citep[e.g.,][]{JY07,J+07,BS10a}.
However, this mechanism requires marginally decoupled dust particles, the formation of which 
is already questioned by the radial drift barrier itself.
Other possibilities include the trapping of dust particles in vortices \citep[e.g.,][]{BS95,KH97} 
and at gas pressure maxima \citep[e.g.,][]{KL07,BHD08,SMI10,P+12}.

This study reexamines this problem by considering a new physical effect: 
porosity evolution of dust aggregates.
Most previous coagulation models \citep[e.g.,][]{NNH81,THI05,BDH08,BDB10}
assumed that dust particles grow with a fixed internal density. 
In reality, however, the internal density of aggregates does change upon collision 
depending on the impact energy.
The evolution of porosity directly affects the growth history of the aggregates 
since the porosity determines the coupling of them to the gas motion.
For example, \citet{OST07} and \citet{Z+11} simulated dust growth with porosity evolution at fixed disk orbital radii and found that  porous evolution delays 
the settling of dust onto the disk midplane. 
However, how the porosity evolution affects the radial drift barrier
has been unaddressed so far.

It has been studied over last two decades how the internal structure changes upon collision 
by laboratory \citep[e.g.,][]{BW00,WGBB09} and numerical 
\citep[e.g.,][]{DT97,W+08,SWT08,SWT12} collision experiments.
One robust finding of these studies is that aggregates grow into low-density, fractal objects 
if the impact energy is lower than a threshold $E_{\rm roll}$ determined by material properties
\citep{BW00,SWT08,OTS09}.
The fractal dimension $d_f$ of the resulting aggregates depends weakly 
on the size ratio between targets and projectiles, 
and falls below two when the target and projectile
have similar sizes \citep{M+92,OTS09}. 
The fractal dimension of two is equivalent to an internal density decreasing 
inversely proportional to the aggregate radius.
The density decrease occurs because each 
merger event involves the creation
of ``voids'' whose volume is comparable to those 
of the aggregates before merger \citep{OTS09}.
\citet{SWT08} estimated the collision energy of aggregates in protoplanetary disks as 
a function of size, and showed that aggregates composed of $0.1\micron$ sized particles 
undergo fractal growth in planet-forming regions until their size reaches centimeters.
This means that the building blocks of planetesimals should have once evolved into 
very fluffy objects with mean internal densities many orders of magnitude lower
than the solid material density. 

More strikingly, recent $N$-body experiments suggest that the porosity of aggregates 
can be kept considerably high even after the collision energy exceeds 
the threshold $E_{\rm roll}$.
\citet{W+08} numerically simulated head-on collisions between equal-sized fractal aggregates 
of $d_f \approx 2$ and found that the fractal dimension after the collision does not exceed $2.5$ 
even at high collision energies.
\citet{SWT08} confirmed this by repeating head-on collisions of the resulting aggregates
at fixed collision velocities.
Furthermore, compaction is even less efficient in offset collisions, where 
the collision energy is spent for stretching rather than compaction of the merged aggregate
\citep{W+07,PD09}.
These results mean that the creation of voids upon merger is nonnegligible 
even when the impact energy is large; in other words, the voids are only imperfectly 
crushed in collisional compaction.
Because of technical difficulty, these theoretical predictions 
have not yet been well tested by laboratory nor microgravity experiments.
Nevertheless, it is worth investigating how aggregates grow and drift inward 
if they evolve into highly porous objects.

In this study, we simulate the temporal evolution of the radial size distribution of aggregates
using the advection--coagulation model developed by \citet{BDH08}. 
Unlike the previous work, we allow the porosities of aggregates 
to change upon collision, depending on their impact energies.
To do so, we adopt the ``volume-averaging method'' proposed by \citet{OTS09}. 
In this method, aggregates of equal mass are regarded as 
having the same volume (or equivalently, the same internal density) 
at each orbital distance, 
and the advection--coagulation equation for the averaged volume 
is solved simultaneously with that for the radial size distribution.
To determine the porosity change upon collisional sticking, we use an analytic recipe  
presented by \citet{SWT12} that well reproduces the collision outcomes 
of recent $N$-body simulations \citep{W+08,SWT08} as a function of the impact energy.
These theoretical tools allow us to study for the first time 
how the porosity evolution affects the growth and radial drift of 
dust aggregates in protoplanetary disks.

In order to clarify the role of porosity evolution, we ignore many other 
effects relevant to aggregate collision, including Coulomb interaction 
\citep{O09,OTTS11a,OTTS11b,MLH12}, bouncing \citep{Z+10,Z+11,W+12}, 
and collisional fragmentation \citep{BDH08,BHD08,BDB09,BKE12}.
Coulomb repulsion due to negative charging can significantly slow down 
the initial fractal growth, but may be negligible once the collisional compaction 
becomes effective \citep{OTTS11b}.
Bouncing is often observed in laboratory experiments for
relatively compact (filling factor $\ga 0.1$) aggregates, 
but is less likely to occur when aggregates are highly porous 
as we consider in this study \citep{LTB08,W+11}.
Seemingly more problematic is fragmentation at high-speed collisions.
This is particularly so when the aggregates are mainly composed of silicate particles,
for which catastrophic disruption begins at collision speeds 
as low as a few ${\rm m~s^{-1}}$ \citep{BW08,W+09,G+10}.
By contrast, collisional fragmentation may be less problematic 
for aggregates made of icy particles, 
for which a higher sticking threshold has been anticipated \citep{CTH93,DT97,G+11}.
For instance, $N$-body collision experiments by \citet{W+09} suggest that 
aggregates made of $0.1~\micron$ sized icy grains do not experience catastrophic disruption
at collision velocities up to $35$--$70~{\rm m~s^{-1}}$.
For this reason, instead of neglecting collisional fragmentation, 
we focus on dust evolution {\it outside the snow line} 
in protoplanetary disks.
A more comprehensive model including the above mentioned effects 
will be presented in future work.

We will show that dust particles evolve into highly porous aggregates
even if collisional compaction is taken into account.
Furthermore, we will show that the porosity evolution
triggers significant acceleration in collisional growth at early stages, 
allowing them to grow across the radial drift barrier in inner regions 
of protoplanetary disks.
Interestingly, this acceleration involves neither enhancement of the collision
velocity nor suppression of the radial drift speed of marginally decoupled
aggregates.
As we will see, this acceleration is a natural consequence of particles' 
aerodynamical property at low Knudsen numbers, i.e., at particle radii 
larger than the mean free path of the gas molecules, 
and the porosity evolution only allows the dust aggregates 
to reach that stage with small aggregate masses. 
Our model calculation shows that the breakthrough 
of the radial drift barrier can occur in ``planet-forming'' regions, 
i.e., at $<10~\AU$ from the central star.
This result suggests that, if the fragmentation of icy aggregates is truly negligible,
the formation of icy planetesimals is possible 
via direct collisional growth of dust particles even without an enhancement of the initial dust-to-gas mass ratio.

This paper is organized as follows. 
In Section~\ref{sec:model}, we describe the disk and collision models that we use in this study.
Simulation results are presented in Section~\ref{sec:results}, 
which we interpret in terms of the timescales for collisional growth 
and radial inward drift in Section~\ref{sec:cond}.
The validity and limitations of our model are discussed in Section~\ref{sec:discussion},
and our conclusions are presented in Section~\ref{sec:summary}.
 
\section{Model}\label{sec:model}
\subsection{Disk Model}\label{sec:disk}
We adopt the minimum-mass solar nebula (MMSN) model of \citet{H81}
with a solar-mass central star.
The radial profiles of the gas surface density $\Sigma_g$ 
and disk temperature $T$ are given by 
$\Sigma_g = 152(r/5~\AU)^{-3/2}~{\rm g~cm^{-2}}$ 
and $T = 125(r/5~\AU)^{-1/2}~{\rm K}$, respectively, 
where $r$ is the distance from the central star.
In this study, we focus on dust evolution outside the snow line, 
which is located at $r \approx 3~\AU$ in the adopted disk model.
The vertical structure is assumed to be in hydrostatic equilibrium,
and hence the vertical structure of the gas density $\rho_g$ is given by
$\rho_g = (\Sigma_g/\sqrt{2\pi}h_g)\exp(-z^2/2h_g^2)$, 
where $h_g = c_s/\Omega$ is the gas scale height, 
$c_s$ is the isothermal sound speed,
and $\Omega$ is the Kepler frequency.
The isothermal sound speed is given by $c_s = \sqrt{\kB T/m_g}$,
where $\kB$ is the Boltzmann constant and $m_g$ is the mean molecular mass.
We assume the mean molecular weight of 2.34, 
which gives $m_g = 3.9\times 10^{-24}~{\rm g}$ and 
$c_s = 6.7\times 10^4(r/5~\AU)^{-1/4}~{\rm cm~s^{-1}}$.
The assumed stellar mass ($1~M_\sun$) 
leads to $\Omega = \sqrt{{\sc G}M_\sun/r^3} 
= 1.8\times 10^{-8}(r/5~\AU)^{-3/2}~{\rm rad~s^{-1}}$ 
and $h_g/r = 0.05 (r/5~\AU)^{1/4}$, 
where $G$ is the gravitational constant.

In reality, protoplanetary disks can be heavier than the MMSN. 
The gravitational stability criterion \citep{T64}
$\Sigma_g < \Omega c_s/\pi G \approx 5.6\times 10^3(r/5~\AU)^{-7/4}~{\rm g~cm^{-2}}$ 
allows the surface density to be up to about 10 times higher than the MMSN value. 
The dependence of our result on the disk mass will be analytically discussed in
Section~\ref{sec:cond}.

Initial dust particles are modeled as compact spheres of 
equal size $a_0 = 0.1~\micron$ and 
equal internal density $\rho_0 = 1.4~{\rm g~cm^{-3}}$,
distributed in the disk with a constant dust-to-gas surface density ratio 
$\Sigma_d/\Sigma_g = 0.01$.
The mass of each initial particle is $m_0 = (4\pi/3)\rho_0 a_0^3 = 5.9\times 10^{-15}~{\rm g}$.
In the following, we will refer to the initial dust particles as ``monomers.''
We define the radius of a porous aggregate as 
$a = [(5/6N)\sum_{i=1}^N\sum_{j=1}^N({\bm x}_i-{\bm x}_j)^2]^{1/2}$, 
where $N$ is the number of the constituent monomers and  
${\bm x}_k~(k=1,2,\dots,N)$ is the position of the monomers \citep{M+92}.
This definition is in accordance with previous $N$-body experiments 
\citep{W+08,SWT08,OTS09} which our porosity model is based on 
(see Section~\ref{sec:porosity}).

Disk turbulence affects the collision and sedimentation of dust particles.
To include these effects, we consider gas turbulence in which the turnover time 
and mean-squared random velocity of the largest turbulent eddies are given by 
$t_L = \Omega^{-1}$ and $\delta v_g^2 = \alpha_D c_s^2$, respectively,
where $\alpha_D$ is the dimensionless parameter characterizing the strength of the turbulence.
The assumption for $t_L$ is based on theoretical anticipation 
for turbulence in Keplerian disks \citep{DV92,FP06,JKM06}. 
The diffusion coefficient for the gas is given by 
$D_g = \delta v_g^2 t_L = \alpha_D c_s^2/\Omega$.
If the gas diffusion coefficient is of the same order as the turbulent viscosity, 
$\alpha_D$ is equivalent to the so-called alpha parameter of \citet{SS73}.
However, we do not consider the viscous evolution of the gas disk for simplicity.
We adopt $\alpha_D = 10^{-3}$ in our numerical simulations.
A higher value of $\alpha_D$ would cause catastrophic collisional fragmentation of aggregates,
which is not considered in this study (see Section~\ref{sec:frag}).

\subsection{Evolutionary Equations}\label{sec:evol}
We solve the evolution of the radial size distribution of dust aggregates 
using the method developed by \citet{BDH08}.
This method assumes the balance between sedimentation
 and turbulent diffusion of aggregates in the vertical direction.
Thus, the vertical number density distribution of aggregates is given  
by a Gaussian $(\calN/\sqrt{2\pi}h_d)\exp(-z^2/2h_d^2)$, 
where $\calN(r,m)$ is  the column number density of aggregates per unit mass 
and $h_d(r,m)$ is the scale height of aggregates at orbital radius $r$ and with mass $m$ \citep{DMS95}.
This approach is valid if the coagulation timescale is longer than 
the settling/diffusion timescale, which is true except for very tiny particles
with short collision times \citep{Z+11}.

The evolution of the radial size distribution $\calN(r,m)$ is given by 
the vertically integrated advection--coagulation equation, which reads \citep{BDH08}
\beqn
\frac{\pd \calN(r,m)}{\pd t} &=& \frac{1}{2}\int_0^m K(r,m',m-m')\calN(r,m')\calN(r,m-m')dm'
 \nonumber \\
&& -  \calN(r,m)\int_0^\infty K(r,m,m')\calN(r,m')dm' \nonumber \\
&& - \frac{1}{r}\frac{\pd}{\pd r}\left[ rv_r(r,m)\calN(r,m)\right],
\label{eq:evolN}
\eeqn
where $v_r$ is the radial drift velocity
and $K$ is the vertically integrated collision rate coefficient given by
\beq
K(r,m_1,m_2) = \frac{\sigma_{\rm coll}}{2\pi h_{d,1} h_{d,2}} \int_{-\infty}^\infty \Delta v
\exp\left(-\frac{z^2}{2h_{d,12}^2}\right)dz.
\eeq
Here, $\sigma_{\rm coll} $ is the collisional cross section, 
$h_{d,1}$ and $h_{d,2}$ are the scale heights of the colliding aggregates 1 and 2, 
$\Delta v$ is the collision velocity, 
and $h_{d,12}= (h_{d,1}^{-2}+h_{d,2}^{-2})^{-1/2}$.
As mentioned in Section~\ref{sec:intro}, we neglect electrostatic and gravitational interactions 
between colliding aggregates and assume perfect sticking upon collision. 
Thus, the collisional cross section is simply given by $\sigma_{\rm coll} = \pi (a_1+a_2)^2$,
where $a_1$ and $a_2$ are the radii of the colliding aggregates.
The validity of neglecting fragmentation will be discussed in Section~\ref{sec:frag}.

The dust scale height $h_d$ in sedimentation--diffusion equilibrium has been analytically obtained 
by \citet{YL07}. For turbulence of $t_L = \Omega^{-1}$ and $D_g = \alpha_D c_s^2/\Omega$, 
it is given by 
\beq
h_d = h_g\left(1+\frac{\Omega t_s}{\alpha_D}\frac{1+2\Omega t_s}{1+\Omega t_s}
\right)^{-1/2},
\label{eq:hd}
\eeq
where $ t_s$ is the stopping time of the aggregates. 
We use this expression in this study.

For the stopping time, we use
\beq
 t_s = \left\{ \begin{array}{ll}
 t_s^{\rm(Ep)} \equiv \dfrac{3m}{4\rho_g v_{\rm th} A}, 
 & a< \dfrac{9}{4}\lambda_{\rm mfp}, \\[3mm]
 t_s^{\rm(St)} \equiv \dfrac{4a}{9\lambda_{\rm mfp}} t_s^{\rm(Ep)},
 & a> \dfrac{9}{4}\lambda_{\rm mfp}, 
\end{array} \right.
\label{eq:ts}
\eeq
where $v_{\rm th} = \sqrt{8/\pi}c_s$ and $\lambda_{\rm mfp}$ are 
the thermal velocity and mean free path of gas particles, respectively, 
and $A$ is the projected area of the aggregate.
The mean free path is related to the gas density as 
\beq
\lambda_{\rm mfp} = \frac{m_g}{\sigma_{\rm mol}\rho_g},
\label{eq:lambda}
\eeq 
where $\sigma_{\rm mol} = 2\times 10^{-15}~{\rm cm^2}$ is 
the collisional cross section of gas molecules.
Our gas disk model gives $\lambda_{\rm mfp} = 120(r/5~\AU)^{11/4}~{\rm cm}$ at the midplane.
Equation~\eqref{eq:ts} satisfies the requirement that the stopping 
time must obey Epstein's law $ t_s  =  t_s^{\rm(Ep)}$ at $a\ll \lambda_{\rm mfp}$ 
and Stokes' law $ t_s  =  t_s^{\rm(St)}$ at $a\gg \lambda_{\rm mfp}$, respectively.
Since $t_s^{\rm(St)} \propto a t_s^{\rm(Ep)}$, 
an aggregate growing in the Stokes regime decouples 
from the gas motion more quickly than in the Epstein regime.
In reality, Stokes' law breaks down when the particle Reynolds number 
(the Reynolds number of flow around the particle) is much greater than unity, 
but we neglect this in our simulations for simplicity.
We will discuss this point further in Section~\ref{sec:Newton}.

The radial drift velocity is taken as 
\beq
v_r = - \frac{2\Omega t_s}{1+(\Omega t_s)^2}\eta v_K,
\label{eq:vr}
\eeq 
where 
\beq
2\eta \equiv -\pfrac{c_s}{v_K}^2 \frac{\pd \ln(\rho_g c_s^2)}{\pd \ln r}
\label{eq:eta}
\eeq
is the ratio of the pressure gradient force to the stellar gravity in the radial direction
and $v_K = r\Omega$ is the Kepler velocity \citep{AHN76,W77,NSH86}.
The radial drift speed has a maximum $\eta v_{\rm K}$, which is realized when $\Omega t_s = 1$. 
In our disk model, $\eta$ scales with $r$ as $\eta = 4.0\times 10^{-3}(r/5~\AU)^{1/2}$,
and the maximum inward speed $\eta v_K = 54~{\rm m~s^{-1}}$ is independent of $r$.
Since $\eta$ is proportional to the gas temperature, 
the maximum drift speed would be somewhat lower in colder disk models \citep{KNH70,HT11}.
Equation~\eqref{eq:vr} neglects the frictional backreaction from dust to gas assuming that 
the local dust-to-gas mass ratio is much lower than unity 
or the stopping time of aggregates dominating the dust mass is much longer than $\Omega^{-1}$.
We examine the validity of this assumption in Section~\ref{sec:NSH}.

In this paper, we also consider the collisional evolution of aggregate porosities. 
We treat the mean volume $V = (4\pi/3)a^3$ of aggregates with orbital 
radius $r$ and mass $m$ as a time-dependent quantity. 
The evolutionary equation for $V(r,m)$ is given by 
\beqn
\frac{\pd \left(V\calN\right)}{\pd t}
 &=& \frac{1}{2}\int_0^m [V_{1+2}K](r,m',m-m') \nonumber \\
&& \times \calN(r,m')\calN(r,m-m')dm'
 \nonumber \\
&& -V(r,m)\calN(r,m)\int_0^\infty K(r,m,m')\calN(r,m')dm' \nonumber \\
&& - \frac{1}{r}\frac{\pd}{\pd r}[rv_r(r,m)V(r,m)\calN(r,m)],
\label{eq:evolVN}
\eeqn
where 
\beq
[V_{1+2}K](r,m_1,m_2) = \frac{\sigma_{\rm coll}}{2\pi h_{d,1} h_{d,2}} \int_{-\infty}^\infty
V_{1+2}\Delta v
\exp\left(-\frac{z^2}{2h_{d,12}^2}\right)dz 
\eeq
with $V_{1+2}$ being the volume of merged aggregates (described in Section~\ref{sec:porosity}).
Equation~\eqref{eq:evolVN} is identical to the original evolutionary equation for $V$ 
derived by \citet[][their Equation~(16)]{OTS09}
except that we here take the vertical integration of the equation 
and take into account the radial advection of dust.
In deriving Equation~\eqref{eq:evolVN}, we have assumed that the dispersion 
of the volume is sufficiently narrow at every $r$ and $m$ \citep[see][]{OTS09}.
This ``volume-averaging'' approximation allows to follow the porosity evolution 
of aggregates without solving higher-order moment equations for the volume, 
and hence with small computational costs. This approximation is valid unless the 
porosity distribution at fixed $r$ and $m$ is significantly broadened by, 
e.g., collisional fragmentation cascades \citep{OTS09}.

\subsection{Dust Model}\label{sec:dust}

\subsubsection{Porosity Change Recipe}\label{sec:porosity}
\begin{figure}[t]
\epsscale{1.1}
\plotone{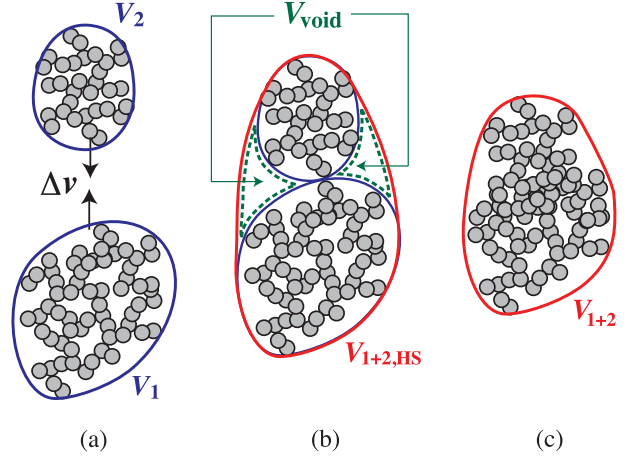}
\caption{Schematic illustration of our porosity change model.
(a) Porous aggregates with volumes $V_1$ and $V_2$ before contact.
(b) Just after contact. At this moment, the volume of the new aggregate is given by 
$V_{1+2,{\rm HS}} = V_1+V_2+V_{\rm void}$, 
where $V_{\rm void}=V_{\rm void}(V_1,V_2)$ is the volume of newly formed voids
(Equation~\eqref{eq:Vvoid}).
 If the collision energy $E_{\rm imp}$ is much smaller than the rolling energy $E_{\rm roll}$, 
the final volume of the new aggregate is equal to $V_{1+2,{\rm HS}}$.
(c) If $E_{\rm imp} \ga E_{\rm roll}$, collisional compression occurs.
In this case, the final volume $V_{1+2} (< V_{1+2,{\rm HS}})$ 
depends on $E_{\rm imp}$.
}
\label{fig:compress}
\end{figure}
The functional form of $V_{1+2}$ determines the evolution of aggregate porosities
in our simulation.
In this study, we give $V_{1+2}$ as a function of the volumes of the colliding aggregates,
$V_1=V(r,m_1)$ and $V_2=V(r,m_2)$, 
and the impact energy $E_{\rm imp} = m_1m_2\Delta v^2/[2(m_1+m_2)]$.
Before introducing the final form of our porosity change recipe (Equation~\eqref{eq:V12}),
we briefly review recent $N$-body collision experiments which our recipe is based on.

Collisional compression depends on the ratio between $E_{\rm imp}$
and the ``rolling energy'' $E_{\rm roll}$ \citep{DT97,BW00,W+07}. 
The rolling energy is defined 
as the energy needed for one monomer to roll over $90^\circ$  
on the surface of another monomer in contact \citep{DT97}.
When $E_{\rm imp} \ll E_{\rm roll}$, two aggregates stick without visible restructuring
(so-called hit-and-stick collision; see Figure~\ref{fig:compress}(b)).
In this case, the volume of the merged aggregate is determined in a geometric manner, 
i.e., independently of $E_{\rm imp}$.
When $E_{\rm imp} \ga E_{\rm roll}$, internal restructuring occurs through inelastic 
rolling among constituent monomers (\citealt*{DT97}; see also Figure~\ref{fig:compress}(c)).
In this case, the final volume $V_{1+2}$ depends on $E_{\rm imp}$ 
as well as on $V_1$ and $V_2$.

For hit-and-stick collisions ($E_{\rm imp}/E_{\rm roll} \to 0$), 
\citet{OTS09} obtained an empirical formula for $V_{1+2}$,
\beq
V_{1+2} = V_{1+2,{\rm HS}} \equiv V_1 + V_2 + V_{\rm void},
\label{eq:V12_HS}
\eeq 
where $V_1$ and $V_2(\leq V_1)$ are the volumes of the two colliding aggregates, and 
\beq
V_{\rm void} = \min \left\{0.99 - 1.03 \ln\pfrac{2}{V_1/V_2+1}, 6.94 \right\}V_2
\label{eq:Vvoid}
\eeq
is the volume of the voids created in the collision (see Figure~\ref{fig:compress}(b)).
For $V_1 \approx V_2$, the void volume  is approximately equal to $V_1$, 
and hence the volume of the new aggregate is approximately given by 
$V_{1+2} \approx 3V_1$. This is equivalent to a fractal relation $V \propto m^{3/d_f}$, 
where $d_f \approx 2$ \citep[see Section~4.2.1 of][]{OTS09}.

In the limit of $E_{\rm imp} \gg E_{\rm roll}$ and 
for head-on collision of equal-sized aggregates ($V_1 = V_2$), 
\citet{SWT08} showed that $V_{1+2}$ obeys the relation
\beq
E_{\rm imp} = -\int_{2^{6/5}V_{1}}^{V_{1+2}} P(V) dV.
\label{eq:SWT08}
\eeq
Here, $P \equiv -dE_{\rm imp}/dV$ 
is the dynamic compression strength of the merged aggregate given by \citep{W+08}
\beq
P(V) = 2\pfrac{5}{3}^{6}\frac{b E_{\rm roll}}{V_0}
\pfrac{\rho_{\rm int}(V)}{\rho_0}^{13/3}N_{1+2}^{2/3},
\label{eq:P}
\eeq
where $b= 0.15$ is a dimensionless fitting parameter,
$V_0 = m_0/\rho_0 = (4\pi/3)a_0^3$ is the monomer volume,
$N_{1+2} = 2m_1/m_0$ is the number of monomers contained in the merged aggregate, 
and $\rho_{\rm int} = 2m_1/V$ is the internal density of the merged aggregate.
If we substitute Equation~\eqref{eq:P} into Equation~\eqref{eq:SWT08}, we obtain 
the equation that explicitly gives $V_{1+2}$ as a function of $E_{\rm roll}$ and $V_1$,
\beq
V_{1+2} =\left[
\dfrac{(3/5)^5 E_{\rm imp}}{N_{1+2}^5bE_{\rm roll}V_0^{10/3}}
+\left(2V_1^{5/6}\right)^{-4} 
\right]^{-3/10}.
\label{eq:V12_SWT08}
\eeq
This equation basically expresses the energy balance 
in collisional compression, but some caution is needed in interpreting it.
First, the initial state for the compression is chosen to be $V=2^{6/5}V_1$,
although the volume just after contact is $V=3V_1$ (see above).
This is based on the fact that compaction from $V=3V_1$ to $V=2^{6/5}V_1$ occurs 
through partial compression of the new voids, which requires little energy \citep{SWT08}.
Second, the dynamic compression strength $P$ depends on mass $N_{1+2}$ 
as well as on internal density $\rho_{\rm int}$, 
meaning that $P$ is not an intensive variable.
This is due to the fact that dynamically compressed parts in the merged aggregate
have a fractal structure with the fractal dimension of $2.5$ \citep{W+08}.
In fact, Equations~\eqref{eq:SWT08}--\eqref{eq:V12_SWT08} 
are more naturally described  
in terms of variables in the 2.5-dimensional space, 
$V_f \propto a^{5/2}$, $\rho_f \propto N_{1+2}/V_f$, and $P_f = -dE_{\rm imp}/dV_f$
\citep[see][]{W+08,SWT08}.
An important point here is that aggregates become 
stronger and stronger against dynamic compression as they grow 
because of the $N_{1+2}^{2/3}$ factor in $P$.

Equations~\eqref{eq:V12_HS} and \eqref{eq:V12_SWT08} express how the volume of the 
merged aggregate is determined in the limits of $E_{\rm imp} \ll E_{\rm roll}$ and
$E_{\rm imp} \gg E_{\rm roll}$, respectively.
To properly take into account intermediate cases ($E_{\rm imp} \sim E_{\rm roll}$),
we adopt an updated analytic formula given recently by \citet{SWT12}.
This reads 
\beq
V_{1+2} = \left\{ \begin{array}{l}
\left[ 
\left(1-\dfrac{E_{\rm imp}}{3bE_{\rm roll}}\right)V_{1+2,{\rm HS}}^{5/6}
 + \dfrac{E_{\rm imp}}{3bE_{\rm roll}}\left(V_{1}^{5/6}+V_{2}^{5/6}\right)
\right]^{6/5} \\[3mm]
\qquad 
({\rm if}~V_{1+2,{\rm HS}}^{5/6}>V_1^{5/6}+V_2^{5/6}~{\rm and}~E_{\rm imp} < 3bE_{\rm roll}), \\[3mm]
\left[
\dfrac{(3/5)^5(E_{\rm imp}-3bE_{\rm roll})}
{N_{1+2}^5bE_{\rm roll}V_0^{10/3}}
+\left(V_1^{5/6}+V_2^{5/6}\right)^{-4}
\right]^{-3/10} \\[4mm]
\qquad
({\rm if}~V_{1+2,{\rm HS}}^{5/6}>V_1^{5/6}+V_2^{5/6}~{\rm and}~E_{\rm imp} > 3bE_{\rm roll}), \\[3mm]
\left[
\dfrac{(3/5)^5 E_{\rm imp}}
{N_{1+2}^5bE_{\rm roll}V_0^{10/3}}
+V_{1+2,{\rm HS}}^{-10/3}
\right]^{-3/10} \\[4mm]
\qquad ({\rm if}~V_{1+2,{\rm HS}}^{5/6}<V_1^{5/6}+V_2^{5/6}), \\
\end{array}
\right.
\label{eq:V12}
\eeq
where $N_{1+2}$ is now defined as $(m_1+m_2)/m_0$.
Note that this equation reduces to Equations~\eqref{eq:V12_HS} 
when $E_{\rm imp} \ll E_{\rm roll}$, and to Equation~\eqref{eq:V12_SWT08} 
when $V_1 = V_2$ and $E_{\rm imp} \gg E_{\rm roll}$.
\citet{SWT12} derived Equation~\eqref{eq:V12} by 
taking into account small energy loss in the partial compression of the new voids.
In addition, unlike Equation~\eqref{eq:V12_SWT08}, 
Equation~\eqref{eq:V12} takes into account the cases where colliding aggregates 
have different volumes and masses ($V_1 \not= V_2$, $m_1 \not = m_2$).
\citet{SWT12} confirmed that 
Equation~\eqref{eq:V12} reproduces the results of numerical collision experiments 
within an error of 20\% as long as the mass ratio $m_2/m_1 (\leq 1)$ 
between the colliding aggregates is larger than 1/16.

We comment on three important caveats regarding our porosity recipe.
First, Equation~\eqref{eq:V12} is still untested for cases where 
colliding aggregates have very different sizes ($m_2/m_1 < 1/16$).
Therefore, the validity of using Equation~\eqref{eq:V12} is at present only guaranteed 
for the case where ``similar-sized'' ($m_2/m_1 \ga 1/16$) 
collisions dominate the growth of aggregates.
We will carefully check this validity in Section~\ref{sec:porous}.
Second, Equation~\eqref{eq:V12} ignores offset collisions, in which 
a considerable fraction of the  impact energy is 
spent for stretching rather than compaction \citep{W+07,PD09}.
For this reason, Equation~\eqref{eq:V12} underestimates 
the porosity increase upon collision. 
Third, we do not consider non-collisional compression 
(e.g., static compression due to gas drag forces), 
which could contribute to the compaction of very large aggregates.
We will discuss the second and third points in more detail in Section~\ref{sec:porosity_validity}. 

In addition to $V$, we need to know the projected area $A$ of aggregates 
to calculate the stopping time $ t_s$.
Unfortunately, a naive relation $A = \pi a^2$ breaks down when the fractal dimension 
of the aggregate is less than 2, since $\pi a^2$ increases faster than mass for this case 
while $A$ does not.
A projected area growing faster than mass means a coupling to the gas becoming 
stronger and stronger as the aggregate grows, which is clearly unrealistic.
To avoid this, we use an empirical formula by \citet{OTS09} that well reproduces 
the mean projected area $\ovl{A}$ of aggregates with monomer number $N=m/m_0$ and radius $a$ 
for both fractal and compact aggregates.
With this formula, all aggregates in our simulations 
are guaranteed to decouple from the gas as they grow.
We remark that this treatment is only relevant to fractal aggregates with $d_f\la 2$; 
for more compact aggregates, the empirical formula reduces to the usual relation 
$A \approx \pi a^2$.

The rolling energy $E_{\rm roll}$ has not been measured so far 
for submicron-sized icy particles, but can be estimated in the following way.
It is anticipated by microscopic friction theory \citep{DT95} that the critical rolling force
$F_{\rm roll} \equiv E_{\rm roll}/(\pi a_0/2)$ is a material constant
(i.e., $E_{\rm roll}$ is proportional to the monomer radius $a_0$).
Recently, a rolling force of $F_{\rm roll} = (1.15\pm 0.24) \times 10^{-3}~{\rm dyn}$ 
has been measured for micron-sized ice particles \citep{G+11}.
Given that $F_{\rm roll}$ is independent of $a_0$, 
the measured force implies the rolling energy 
of $E_{\rm roll} = (\pi a_0/2)F_{\rm roll}
\approx 1.8 \times 10^{-8}~{\rm erg}$ for $a_0 = 0.1\micron$.
We use this value in our simulations.

\subsubsection{Collision Velocity}\label{sec:dv}
We consider Brownian motion,  radial and azimuthal drift, vertical settling, and turbulence 
as sources of the collision velocity,
and give the collision velocity $\Delta v$ as the root sum square of these contributions, 
\beq
\Delta v = \sqrt{(\Delta v_B)^2+(\Delta v_r)^2+(\Delta v_\phi)^2+(\Delta v_z)^2+(\Delta v_t)^2},
\eeq
where $\Delta v_B$, $\Delta v_r$, $\Delta v_\phi$, $\Delta v_z$, and $\Delta v_t$
are the relative velocities induced by Brownian motion, radial drift, azimuthal drift, vertical settling,
and turbulence, respectively.

The  Brownian-motion-induced velocity is given by
\beq
\Delta v_B = \sqrt{\frac{\pi m_1m_2}{8(m_1+m_2)k_{\rm B}T}},
\label{eq:vB}
\eeq
where $m_1$ and $m_2$ are the masses of the two colliding aggregates.

The relative velocity due to radial drift is given by 
$\Delta v_r = |v_r(t_{s,1}) - v_r(t_{s,2})|$,
where $t_{s,1}$ and $t_{s,2}$ are the stopping times of the colliding aggregates,
and $v_r$ is the radial velocity given by Equation~\eqref{eq:vr}.
Similarly, the relative velocity due to differential azimuthal motion is given by 
$\Delta v_\phi = |v'_{\phi}(t_{s,1}) - v'_{\phi}(t_{s,2})|$,
where 
\beq
v'_\phi = -\frac{\eta v_K}{1+(\Omega t_s)^2}
\label{eq:vphi}
\eeq
is the deviation of the azimuthal velocity from the local Kepler velocity \citep{AHN76,W77,NSH86}.
Here, we have neglected the backreaction from dust to gas as we already did for 
the radial velocity (see Sections~\ref{sec:dv} and \ref{sec:NSH}).

For the differential settling velocity, we assume 
$\Delta v_z = |v_z(t_{s,1}) - v_z(t_{s,2})|$, where 
\beq
v_z = - \frac{\Omega^2 t_sz}{1+\Omega t_s}.
\label{eq:vz}
\eeq
Equation~\eqref{eq:vz} reduces to the terminal settling velocity $v_z = -\Omega^2 t_sz$
in the strong coupling limit $\Omega t_s \ll 1$, and to the amplitude of the vertical oscillation 
velocity at $\Omega t_s \gg 1$ \citep{BDH08}.

For the turbulence-driven relative velocity, we use an analytic formula 
for Kolmogorov turbulence derived by \citet[][their Equation~(16)]{OC07}.
This analytic formula has three limiting expressions 
(Equations~(27)--(29) of \citealt*{OC07}):
\beq
\Delta v_t \approx \left\{\begin{array}{ll}
\delta v_g {\rm Re}_t^{1/4}\Omega |t_{s,1} - t_{s,2}|, & t_{s,1} \ll t_\eta, \\[3mm]
(1.4\dots 1.7)\times \delta v_g\sqrt{\Omega t_{s,1}},  & t_\eta \ll t_{s,1} \ll \Omega^{-1}, \\
\delta v_g \left(\dfrac{1}{1+\Omega t_{s,1}}+\dfrac{1}{1+\Omega t_{s,2}} \right)^{1/2},
& \Omega t_{s,1} \gg 1,
\end{array} \right.
\label{eq:vT}
\eeq
where ${\rm Re}_t$ is  the turbulent Reynolds number,
$t_\eta = {\rm Re}_t^{-1/2}t_L$ is the turnover time of the smallest eddies,
and the numerical prefactor $(1.4 \dots 1.7)$ in the second equality 
depends on the ratio between the stopping times, $t_{s,2}/t_{s,1}$.
The turbulent Reynolds number is given by ${\rm Re}_t = D_g/\nu_{\rm mol}$,
where $\nu_{\rm mol} = v_{\rm th}\lambda_{\rm mfp}/2$ is the molecular viscosity.
For $t_{s,1} \sim t_{s,2}$, the maximum induced velocity is $\Delta v_t \approx \delta v_g$, 
which is reached when $\Omega t_{s,1} \approx 1$.

When two colliding aggregates belong to the Epstein regime 
and their stopping times are much shorter than $t_\eta~(\ll \Omega^{-1})$,
the relative velocity driven by sedimentation and turbulence is approximately
proportional to the difference between the mass-to-area ratios $m/A$ of two colliding aggregates.
In this case, as pointed out by \citet{OTTS11a}, the dispersion of the mass-to-area ratio
becomes important for fractal aggregates of $d_f \la 2$, since the mean mass-to-area 
ratio of the aggregates approaches to a constant and hence the difference in $m/\ovl{A}(m)$ 
vanishes.
To take into account the dispersion effect, we evaluate the differential mass-to-area ratio as 
$|\Delta(m/A)|^2 = |m_1/\ovl{A}_1-m_2/\ovl{A}_2|^2 
+ \eps^2[(m_1/\ovl{A}_1)^2+(m_2/\ovl{A}_2)^2]$,
where $\ovl{A}_j = \ovl{A}(m_j)$  $(j=1,2)$
are the mean projected area of aggregates with mass $m_j$ 
(see Section~\ref{sec:porosity}) and 
$\eps$ is the standard deviation of the mass-to-area ratio divided by the mean
\citep[for the derivation, see the Appendix of][]{OTTS11a}.
We assume $\eps = 0.1$ in accordance with the numerical estimate by \citet{OTTS11a}.

\subsection{Numerical Method}\label{sec:method}
We solve Equations~\eqref{eq:evolN} and \eqref{eq:evolVN} numerically with 
an explicit time-integration scheme and a fixed-bin method. 
The radial domain is taken to be outside the snow line, $3~\AU \leq r \leq 150~\AU$, 
discretized into 100 rings with an equal logarithmic width 
$\Delta \ln(r[\AU]) = (\ln 150-\ln 3)/100$.
The advection terms are calculated by the spatially first-order upwind scheme.
We impose the outflow and zero-flux boundary conditions at the innermost and outermost 
radii ($r=3~\AU$ and 150~\AU), respectively; thus, the total dust mass inside the domain is a decreasing function of time. Our numerical results are unaffected 
by the choice of the boundary condition at the outermost radius, since dust growth at this location 
is too slow to cause appreciable radial drift within the calculated time. 
The coagulation terms are calculated by the method given by \citet{OTS09}.
Specifically, at the center of each radial ring we divide the mass coordinate into linearly spaced bins 
$m_k = km_0~(k=1,2,\dots,N_{bd})$ for $m \leq N_{bd}m_0$ and logarithmically spaced bins 
$m_k = m_{k-1}10^{1/N_{bd}}~(k=N_{bd}+1,\dots)$ for $m > N_{bd}m_0$, where 
$N_{bd}$ is an integer. We adopt $N_{bd} = 40$; as shown by \citet{OTS09}, the calculation 
results well converge as long as $N_{bd} \geq 40$.
The time increment $\Delta t$ is adjusted at every time step 
so that the fractional decreases in $\calN$ and $V\calN$ 
fall below 0.5 (i.e., $\Delta t < - 0.5(\pd\ln\calN/\pd t)^{-1}$ and $\Delta t < - 0.5(\pd\ln V\calN/\pd t)^{-1}$)
at all bins.

\section{Results}\label{sec:results}

\subsection{Compact Aggregation}\label{sec:compact}
\begin{figure}
\epsscale{1.1}
\plotone{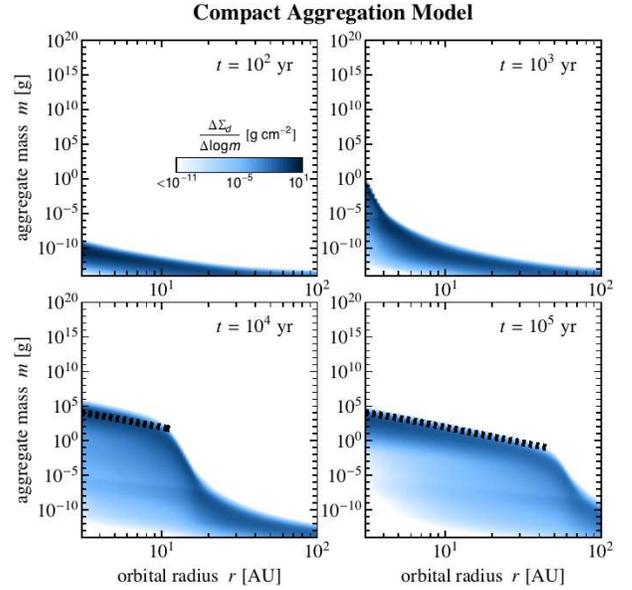}
\caption{Aggregate size distribution $\Delta\Sigma_d/\Delta\log m$ at different times $t$
for the compact aggregation model ($\rho_{\rm int} = 1.4~{\rm g~cm^{-3}}$)
as a function of orbital radius $r$ and aggregate mass $m$. 
The dotted lines mark the aggregate size at which $\Omega t_s$ exceeds $0.1$.
}
\label{fig:evolC}
\end{figure}
\begin{figure}
\epsscale{1.1}
\plotone{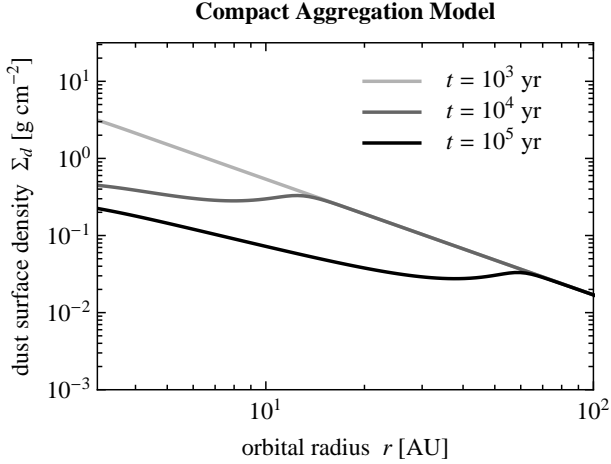}
\caption{Radial profiles of the total dust surface density $\Sigma_d$ at different times 
for the compact aggregation model ($\rho_{\rm int} = 1.4~{\rm g~cm^{-3}}$). 
}
\label{fig:sigmaC}
\end{figure}
To begin with, we show the result of compact aggregation.
In this simulation, we fixed the internal density $\rho_{\rm int} \equiv m/V$ 
of the aggregates to the material density $\rho_0 = 1.4~{\rm g~cm^{-3}}$, and solved only 
the evolutionary equation for the radial size distribution ${\cal N}(r,m)$ (Equation~\eqref{eq:evolN}),
as done in previous studies \citep[e.g.,][]{BDH08}.
Figure~\ref{fig:evolC} shows the snapshots of the radial size distribution at different times.
Here, the distribution is represented by the dust surface density per decade of
aggregate mass, $\Delta\Sigma_d/\Delta\log m \equiv \ln(10) m^2 {\cal N}(r,m)$.
At each orbital radius, dust growth proceeds without significant radial drift
until the stopping time of the aggregates reaches $\Omega  t_s \sim 0.1$ 
(the dotted lines in Figure~\ref{fig:evolC}).
However, as the aggregates grow, the radial drift becomes faster and faster, 
and further growth becomes limited only along the line $\Omega  t_s \sim 0.1$ 
on the $r$--$m$ plane.
Figure~\ref{fig:sigmaC} shows the evolution of the total dust surface density 
$\Sigma_d \equiv \int m{\cal N} dm = \int (\Delta\Sigma_d/\Delta\log m)d\log m$.
We see that a significant amount of dust has been lost from the planet-forming region 
$r\la 30~\AU$ within $10^5~{\rm yr}$.
In this region, the dust surface density scales as $r^{-1},$\footnote{
It can be analytically shown \citep{BKE12} that the dust surface density profile obeys a scaling
$\Sigma_d \propto \sqrt{\Sigma_g/(r^2\Omega)}$ ($\propto r^{-1}$ for 
$\Sigma_g\propto r^{-3/2}$) when radial drift balances with turbulence-driven growth. 
} and hence the dust-to-gas surface density ratio $\propto r^{-1}/\Sigma_g \propto r^{1/2}$ 
decreases toward the central star. 

\begin{figure}
\epsscale{1.1}
\plotone{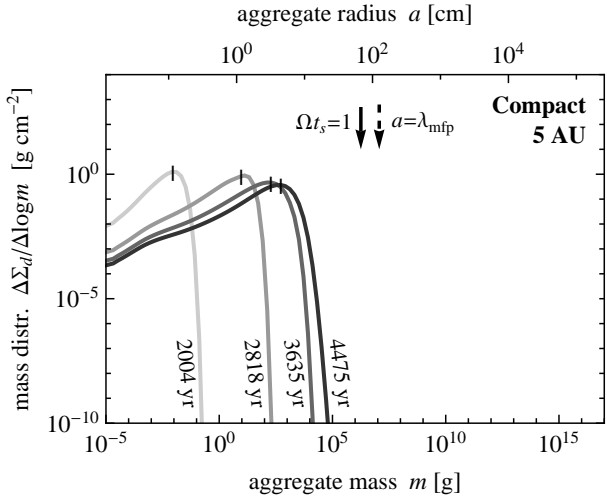}
\caption{Aggregate size distribution $\Delta\Sigma_d/\Delta\log m$ at $r = 5~\AU$ 
and $t = 2000~{\rm yr}$--$4470~{\rm yr}$ for the compact aggregation model. 
The dashed and solid arrows indicate the aggregate sizes at which $a = \lambda_{\rm mfp}$
and $\Omega t_s = 1$, respectively.
Shown at the top of the panel is the aggregate radius $a$. 
The vertical bars indicate the weighted average mass $\bracket{m}_m$ (Equation~\eqref{eq:mavr}).
}
\label{fig:5AU_C}
\end{figure}
Figure~\ref{fig:5AU_C} shows the evolution of the dust size distribution observed at $r=5~\AU$.
Here, in order to characterize the typical aggregate size at each evolutionary stage, 
we introduce the weighted average mass $\bracket{m}_m$ defined by 
\beq
\bracket{m}_m \equiv \frac{\int m^2{\cal N}dm}{\int m{\cal N} dm}
= \frac{1}{\Sigma_d}\int m \frac{\Delta\Sigma_d}{\Delta\log m}d\log m.
\label{eq:mavr}
\eeq
The weighted average mass approximately corresponds to 
the aggregate mass at the peak of 
$\Delta\Sigma_d/\Delta\log m$ \citep[see, e.g.,][]{OST07,OTTS11a}. 
In Figure~\ref{fig:5AU_C}, the weighted average mass at each time is indicated by the short 
vertical line. 
At $r = 5~\AU$, the growth--drift equilibrium is reached 
at $t\approx 4000~{\rm yr}$,
and  the typical size of the aggregates is 
$\bracket{m}_m \approx 500~{\rm g}$ in mass 
($\approx 4$~{\rm cm} in radius, $\approx 0.07\Omega^{-1}$ in stopping time).
Note that the final aggregate radius is much smaller than the mean free path $\lambda_{\rm mfp}$
of gas molecules (the dashed arrow in Figure~\ref{fig:5AU_C}), 
which means that the gas drag onto the aggregates is determined by Epstein's law.
As we will see in the following, porosity evolution allows aggregates 
to reach the Stokes drag regime at much smaller $\Omega t_s$.

\subsection{Porous Aggregation}\label{sec:porous}
\begin{figure*}[t]
\epsscale{1.1}
\plottwo{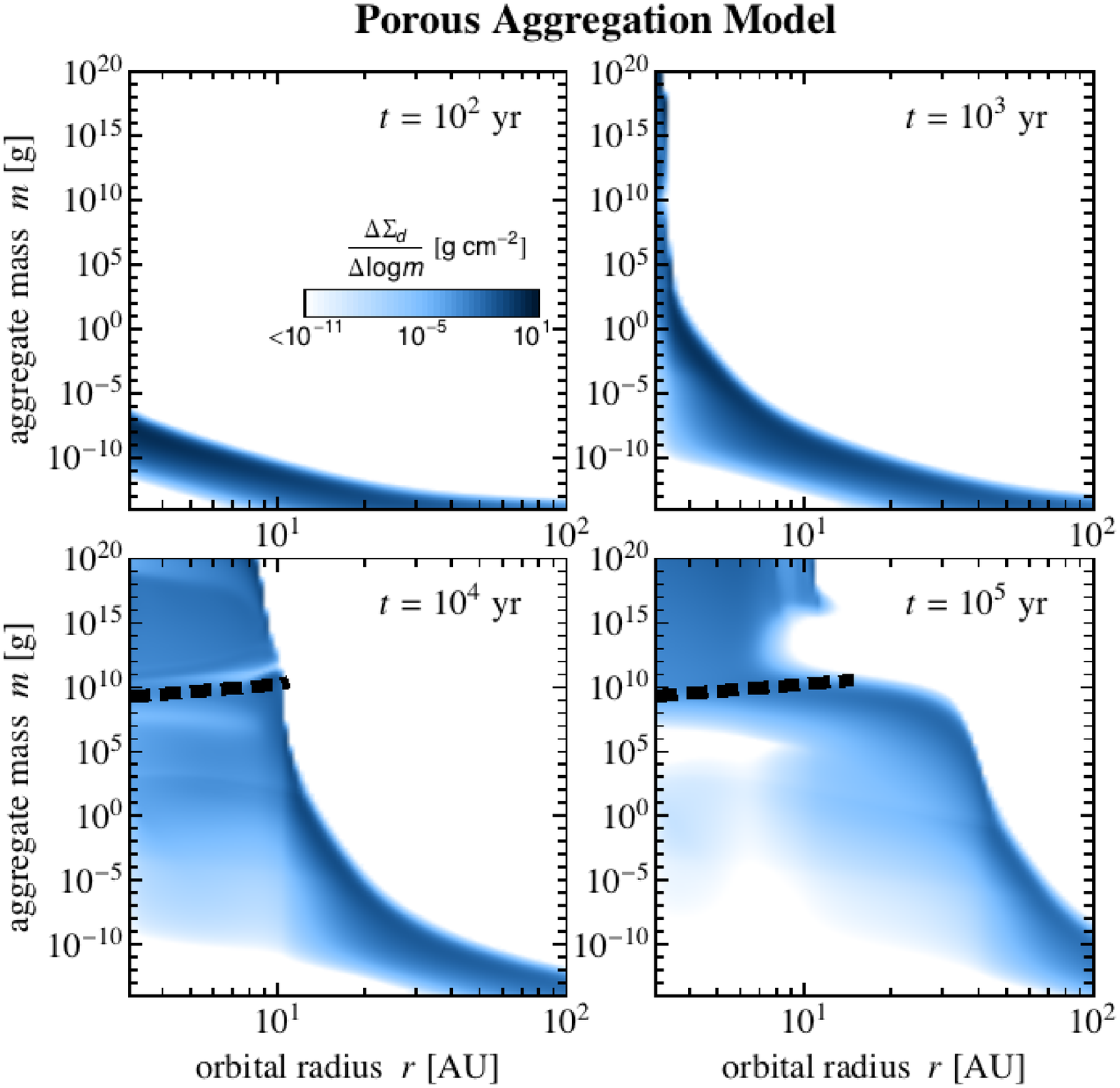}{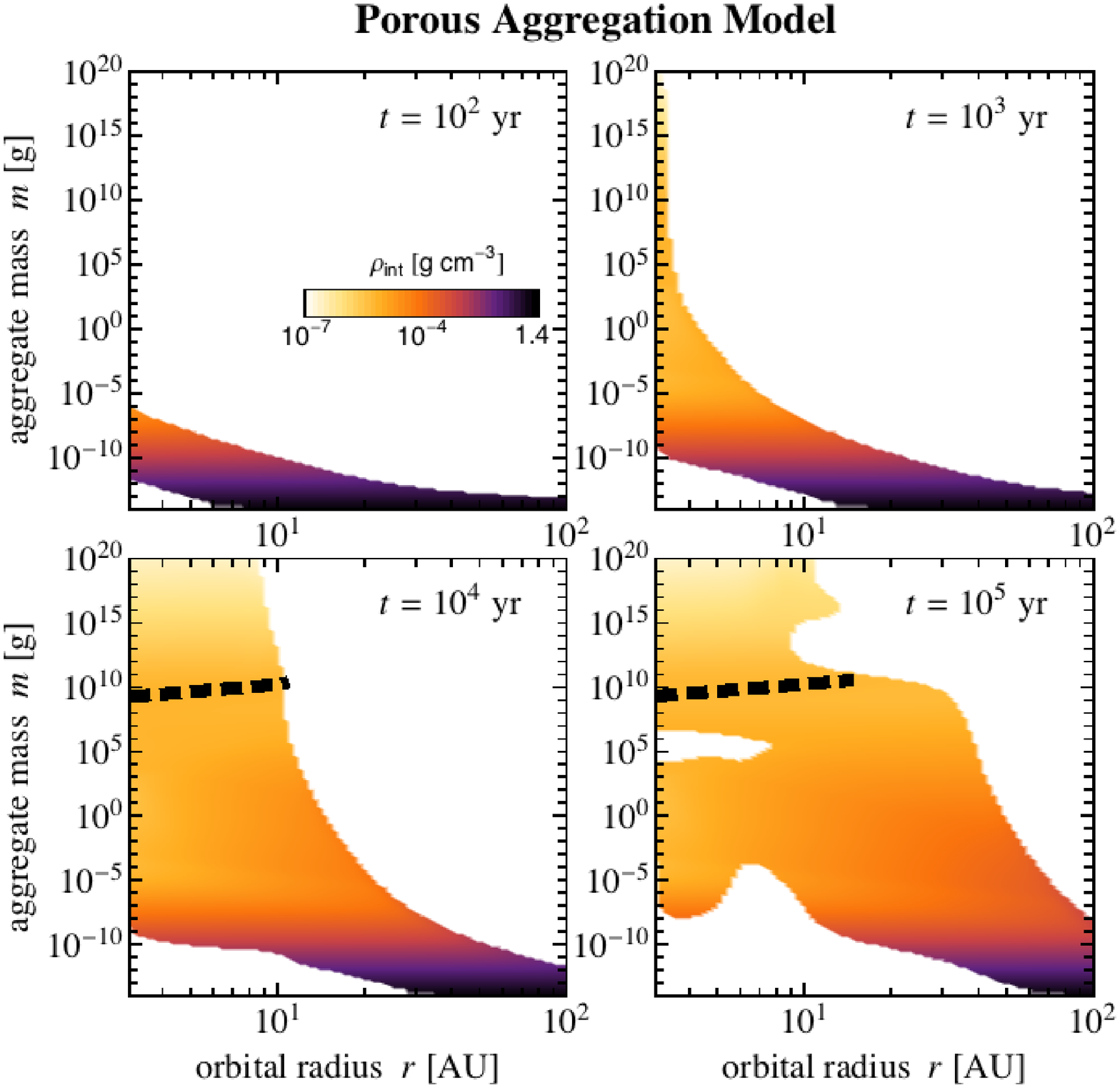}
\caption{Aggregate size distribution $\Delta\Sigma_d/\Delta\log m$ (left four panels) 
and internal density $\rho_{\rm int} = m/V$ (right four panels) 
at different times $t$ for the porous aggregation model as a function of orbital radius $r$ and 
aggregate mass $m$.
The dashed lines mark the aggregate size at which $\Omega t_s$ exceeds unity.
}
\label{fig:evolP}
\end{figure*}

\begin{figure}
\epsscale{1.1}
\plotone{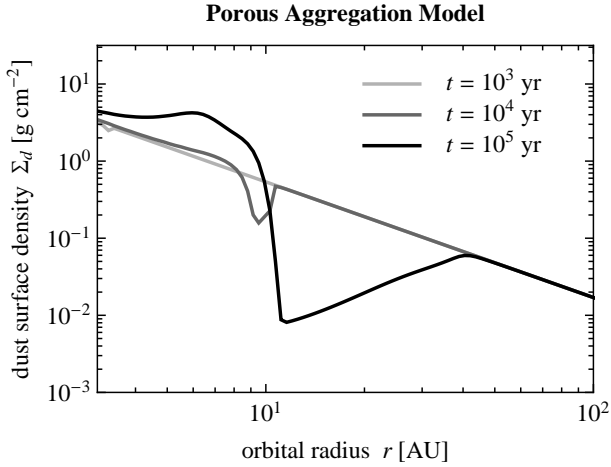}
\caption{Radial profiles of the total dust surface density $\Sigma_d$ at different times 
for the porous aggregation model.
}
\label{fig:sigmaP}
\end{figure}

Now we show how porosity evolution affects dust evolution.
Here, we solve the evolutionary equation for $V(r,m)$ (Equation~\eqref{eq:evolVN})
simultaneously with that for ${\cal N}(r,m)$ (Equation~\eqref{eq:evolN}).
The result is shown in Figure~\ref{fig:evolP}, which displays
 the snapshots of the aggregate size distribution $\Delta\Sigma_d/\Delta\log m$ and internal density $\rho_{\rm int} = m/V$ at different times $t$ as a function of $r$ and $m$.
The evolution of the total dust surface density $\Sigma_d$ is shown in Figure~\ref{fig:sigmaP}.

The left four panels of Figure~\ref{fig:evolP} 
show how the radial size distribution evolves in the porous aggregation.
At $t < 10^3~{\rm yr}$, the evolution is qualitatively similar 
to that in compact aggregation (Section~\ref{sec:compact}). 
However, in later stages, the evolution is significantly different. 
We observe that aggregates in the inner region of the disk ($r < 10~\AU$)
undergo rapid growth and eventually overcome the radial drift barrier lying at $\Omega t_s \sim 1$ 
(dashed lines in Figure~\ref{fig:evolP}) within $t \sim 10^4~{\rm yr}$.
At this stage, the radial profile of the total dust surface $\Sigma_d$ is hardly changed from the initial profile, as is seen in Figure~\ref{fig:sigmaP}.
In the outer region ($r > 10~\AU$), 
aggregates do drift inward before they reach $\Omega t_s \sim 1$ 
as in the compact aggregation model.
However, unlike in the compact aggregation, the inward drift results in the pileup of dust materials in the inner region ($r \approx 4$--$9~\AU$) 
rather than the loss of them from outside the snow line (see Figure~\ref{fig:sigmaP}).
This occurs because most of the drifting aggregates get 
captured by aggregates that have already overcome the drift barrier.
As a result of this, the dust-to-gas mass ratio in the inner regions is enhanced by a factor of several 
in $10^5~{\rm yr}$.

The right four panels of Figure~\ref{fig:evolP} show the evolution 
of the internal density $\rho_{\rm int} = m/V$ as a function of $r$ and $m$.
First thing to note is that the dust particles grow into low-density objects at every location
until their internal density reaches $\rho_{\rm int} \sim 10^{-5}$--$10^{-3}~{\rm g~cm^{-3}}$.
In this stage, the internal density decreases as $\rho_{\rm int} \approx (m/m_0)^{-1/2}\rho_0$,
meaning that the dust particles grow into fractal aggregates with the fractal dimension 
$d_f \approx 2$ \citep{OTS09}. 
The fractal growth generally occurs in early growth stages where 
the impact energy is too low to cause collisional compression, i.e., $E_{\rm imp} \ll E_{\rm roll}$ 
\citep[e.g.,][]{B04,OST07,Z+10}.
At $m \sim 10^{-4}$--$10^{-6}~{\rm g}$, the fractal growth stage terminates, 
followed by the stage where collisional compression becomes effective 
($E_{\rm imp} \gg E_{\rm roll}$).
In this late stage, the internal density decreases more slowly or 
is kept at a constant value depending on the orbital radius.
We will examine the density evolution in more detail in Section~\ref{sec:density}.

\begin{figure}
\epsscale{1.1}
\plotone{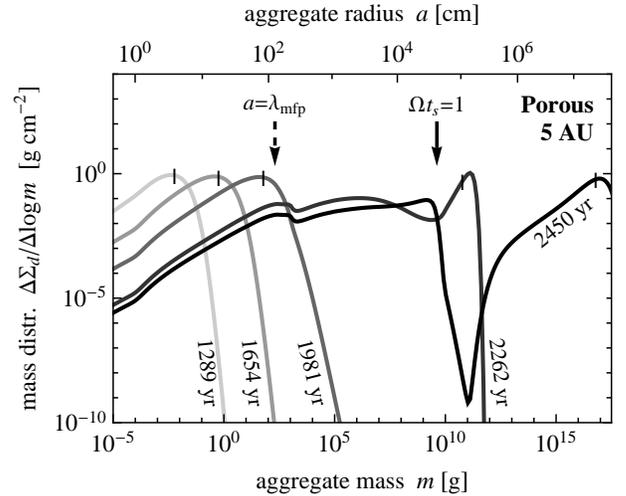}
\caption{Aggregate mass distribution $\Delta\Sigma_d/\Delta\log m$ at $r = 5~\AU$ and
$t = 1289~{\rm yr}$--$2450~{\rm yr}$ for the porous aggregation model.
The dashed and solid arrows indicate the sizes at which $a = \lambda_{\rm mfp}$
and $\Omega t_s = 1$, respectively.
Shown at the top of the panel is the aggregate radius $a$ measured at $t = 2450~{\rm yr}$. 
The vertical bars indicate the weighted average mass $\bracket{m}_m$ (Equation~\eqref{eq:mavr}).
}
\label{fig:5AU}
\end{figure}
\begin{figure}
\epsscale{1.1}
\plotone{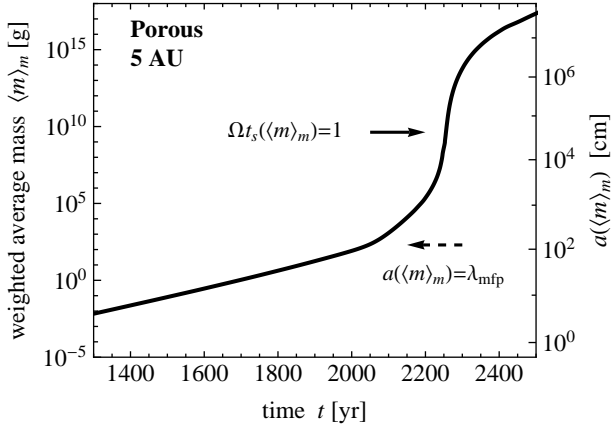}
\caption{Weighted average mass $\bracket{m}_m$ (Equation~\eqref{eq:mavr}) at $r=5~\AU$
as a function of time $t$ for the porous aggregation model.
Shown at the right of the panel is the corresponding aggregate radius $a(\bracket{m}_m)$.  
The dashed and solid arrows indicate the sizes at which $a(\bracket{m}_m) = \lambda_{\rm mfp}$
and $\Omega t_s(\bracket{m}_m) = 1$, respectively.
}
\label{fig:tm}
\end{figure}

Figure~\ref{fig:5AU} shows the evolution of the mass distribution function at $r = 5~\AU$  
during $t \approx 1200$--$2500~{\rm yr}$.
The evolution of the weighted average mass $\bracket{m}_m$ is shown in Figure~\ref{fig:tm}.
It is seen that the acceleration of the growth begins when the aggregate size $a$
exceeds the mean free path of gas molecules, $\lambda_{\rm mfp}$ 
(the dashed arrow in Figure~\ref{fig:5AU}).
This suggests that the acceleration is due to the change in the aerodynamical 
property of the aggregates.
At $a \approx \lambda_{\rm mfp}$, the gas drag onto the aggregates 
begins to obey Stokes' law.
In the Stokes regime, the stopping time $t_s$ of aggregates quickly increases with size
(see Section~\ref{sec:evol}).
This causes the quick growth of the aggregates since
the relative velocity between aggregates increases with $t_s$ (as long as $\Omega t_s <1$). 
As a result of the growth acceleration, the aggregates grow from 
$a\approx \lambda_{\rm mfp}$ to $\Omega t_s \approx 1$ 
within $300~{\rm yr}$, 
which is short enough for them to break through the radial drift barrier.

The decrease in the internal density plays an important role on 
the growth acceleration. 
More precisely, the low internal density allows the aggregates to
reach $a\approx \lambda_{\rm mfp}$ at early growth stages, i.e., at small $\Omega t_s$.
In fact, the growth acceleration was not observed in the compact aggregation,
since the aggregate size is smaller than the mean free path at {\it all} 
$\Omega t_s <1$ (see Figure~\ref{fig:5AU_C}).
A more rigorous explanation for this will be given in Section~\ref{sec:cond}.

\subsubsection{Projectile Mass Distribution}\label{sec:massratio}
As noted in Section~\ref{sec:porosity}, our porosity change model has only been 
tested for collisions between similar-sized aggregates.  
To check the validity of using this model, 
we introduce the projectile mass distribution function \citep{OTS09}:
\beq
C_{m_t}(m_p) \equiv \frac{m_p K(m_p,m_t){\cal N}(m_p)}
{\int_0^{m_t} m_p' K(m_p',m_t){\cal N}(m_p')dm_p'}, \quad m_p \leq m_t,
\label{eq:collrate}
\eeq
which is normalized so that $\int_0^{m_t} C_{m_t}(m_p) dm_p = 1$.
The denominator of $C_{m_t}(m_p)$ is equal to 
the growth rate $t_{\rm grow}^{-1} \equiv d\ln m_t/dt$ 
of a target having mass $m_t$
\citep[see][]{OTS09}. Hence, the quantity $C_{m_t}(m_p) dm_p$ 
measures the contribution of projectiles within mass range $[m_p,m_p+dm_p]$ 
to the growth of the target.

\begin{figure}
\epsscale{1.1}
\plotone{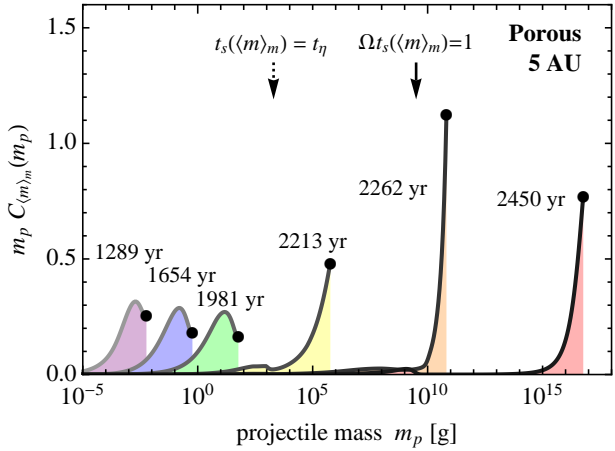}
\caption{Normalized projectile mass distribution per unit logarithmic projectile mass,
$m_pC_{m_t}(m_p)$, for a target with mass $m_t = \bracket{m}_m$ 
at different times $t(=1289~{\rm yr}$--$2450~{\rm yr})$ at $r = 5~\AU$ for the porous aggregation model
(see Equation~\eqref{eq:collrate} for the definition of $C_{m_t}(m_p)$).
The filled circles show the values for equal-sized collisions, $m_p = m_t (=\bracket{m}_m)$.
The dotted and solid arrows indicate the target mass at which $t_s = t_\eta$
and $\Omega t_s = 1$, respectively.
}
\label{fig:rate}
\end{figure}
Figure~\ref{fig:rate} shows the projectile mass distribution per unit $\ln m_p$, 
$m_pC_{m_t}(m_p)$, for targets with mass $m_t = \bracket{m}_m$ at $r = 5~\AU$ and at different $t$.
We see that the growth of the $m_t=\bracket{m}_m$ target is dominated by 
projectiles within a mass range $0.1m_t \la m_p \leq m_t$. 
In fact, the projectile mass distribution integrated over $0.1m_t \leq m_p \leq m_t$
exceeds 50\% for all the cases presented in Figure~\ref{fig:rate}.
This means that the growth of aggregates is indeed dominated by collisions 
with similar-sized ones as required by the limitation of our porosity model.
This is basically the consequence of 
the fact that the aggregate mass distribution $\Delta \Sigma_d/\Delta \log m$ 
is peaked around the target mass $m \sim \bracket{m}_m$
(see Figure~\ref{fig:5AU}).
The mass ratio $m_p/m_t$ at the peak of $m_pC_{m_t}(m_p)$ 
reflects the size dependence of the turbulence-driven relative velocity $\Delta v_t$, 
which is the main source of the collision velocity in our simulation.
At $t \la 2000~{\rm yr}$ ($\bracket{m}_m \la 10^3~{\rm g}$), 
the dominant projectile mass is lower than $m_t(=\bracket{m}_m)$,
since both the target and projectiles are tightly coupled to turbulence 
(i.e., $t_s(m_t), t_s(m_p) < t_\eta$)
and hence $\Delta v_t$ vanishes at equal-sized collisions
(see the first expression of Equation~\eqref{eq:vT}).
At $t \ga 2000~{\rm yr}$ ($\bracket{m}_m \ga 10^3~{\rm yr}$), 
the target decouples from small turbulent eddies ($t_s(m_t) > t_\eta$). 
This results in the shift of the dominant collision mode to $m_p \approx m_t$ 
because $\Delta v_t$ no more vanishes 
at equal-sized collisions (see the second line of Equation~\eqref{eq:vT}).
 
\subsubsection{Density Evolution History}\label{sec:density}
\begin{figure}
\epsscale{1.1}
\plotone{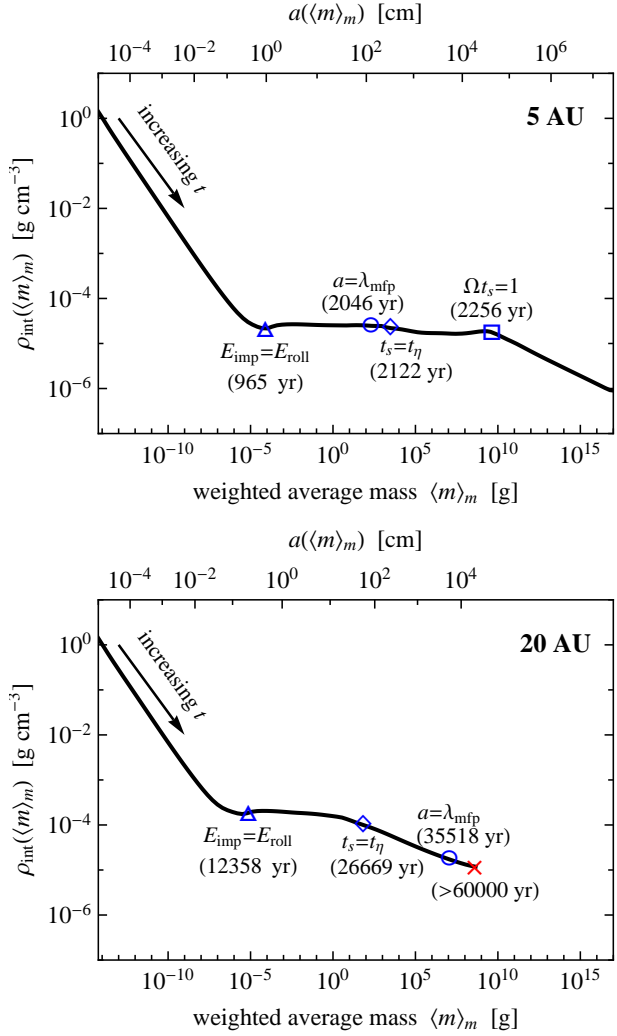}
\caption{Temporal evolution of the weighted average mass $\bracket{m}_m$ 
and the internal density $\rho_{\rm int}(\bracket{m}_m)$ at orbital radii
$r = 5~\AU$ (upper panel) and 20~AU (lower panel).
Shown at the top of the panels is the aggregate radius $a(\bracket{m}_m)$ 
at each orbital radius.
The triangles, circles, diamonds, and square mark the sizes at which 
$E_{\rm imp}=E_{\rm roll}$, $a = \lambda_{\rm mfp}$, $t_s = t_\eta$, 
and $\Omega t_s = 1$, respectively.
At $r = 20~\AU$, dust growth stalls due to the radial drift barrier (cross symbol) 
before reaching $\Omega t_s = 1$.
}
\label{fig:rho}
\end{figure}

To see the density evolution history in detail,
we plot in Figure~\ref{fig:rho} the temporal evolution of the weighted average mass
$\bracket{m}_m$ and the internal density of aggregates with mass $m=\bracket{m}_m$
at orbital radii $r = 5~\AU$ and $20~\AU$. 

As mentioned above, dust particles initially grow into fractal aggregates of $d_f \approx 2$ 
until the impact energy $E_{\rm imp}$ becomes comparable to the rolling energy $E_{\rm roll}$.
With this fact, one can analytically estimate the aggregate size at which the fractal growth terminates.
Our simulation shows that the collision velocity between the fractal aggregates 
is approximately given by the turbulence-driven velocity in the strong-coupling limit
(Equation~\eqref{eq:vT} with $t_s \ll t_\eta$).
Assuming that the collisions mainly occur between aggregates of similar sizes
(see Section~\ref{sec:massratio}),  
the reduced mass and the collision velocity are roughly given by $m/2$ and
$\delta v_g {\rm Re}_t^{1/4}\Omega t_s$, respectively.
In addition, we use the fact that fractal aggregates with $d_f \approx 2$ have 
the mass-to-area ratio comparable to their constituent monomers.
This means that the stopping time of the aggregates is as short as the monomers 
and is hence given by Epstein's law. 
Thus, the impact energy is approximated as 
\beq
E_{\rm imp} \approx \frac{m}{4}\Delta v_t^2 
\approx \frac{3}{8}m \pfrac{\delta v_g {\rm Re}_t^{1/4}\Omega}{\rho_g v_{\rm th}}^2
\pfrac{3m}{4A}^2.
\label{eq:Eimp_frac}
\eeq
Furthermore, using the definitions for $\rho_g$, $v_{\rm th}$, and ${\rm Re}_t$, 
we have $\rho_g v_{\rm th} = (2/\pi)\Sigma_g\Omega$ and 
${\rm Re}_t = \alpha_D\Sigma_g\sigma_{\rm mol}/(2m_g)$ for the midplane.
Substituting them into Equation~\eqref{eq:Eimp_frac} and using $\delta v_g =\sqrt{\alpha_D} c_s$
and $m/A \approx m_0/(\pi a_0^2) = 4\rho_0 a_0/3$, we obtain
\beq
E_{\rm imp} \approx \frac{3\pi^2}{32\sqrt{2}}\alpha_D^{3/2} m c_s^2
\pfrac{\Sigma_g \sigma_{\rm mol}}{m_g}^{1/2}\pfrac{\rho_0a_0}{\Sigma_g}^2.
\label{eq:Eimp_frac2}
\eeq
Thus, the impact energy is proportional to the mass.
We define the rolling mass $m_{\rm roll}$ by the condition $E_{\rm imp} = E_{\rm roll}$.
Using Equation~\eqref{eq:Eimp_frac2}, the rolling mass is evaluated as
\beqn
m_{\rm roll} &\approx& \frac{32\sqrt{2}}{3\pi^2}
\frac{E_{\rm roll}}{c_s^2\alpha_D^{3/2}} \pfrac{m_g}{\Sigma_g \sigma_{\rm mol}}^{1/2}
\pfrac{\Sigma_g}{\rho_0a_0}^2
\nonumber \\
&\sim& 10^{-4}~{\rm g}~\pfrac{\alpha_D}{10^{-3}}^{-3/2}\pfrac{T}{100~{\rm K}}^{-1}
\pfrac{\Sigma_g}{100~{\rm g~cm^{-2}}}^{3/2}
\nonumber \\
&&\times\pfrac{F_{\rm roll}}{10^{-3}~{\rm dyn}}\pfrac{\rho_0}{1~{\rm g~cm^{-3}}}^{-2}
\pfrac{a_0}{0.1~\micron}^{-1},
\label{eq:mroll}
\eeqn
where we have used that $E_{\rm roll} = (\pi a_0/2)F_{\rm roll}$ (see Section~\ref{sec:porosity}).
Using the relations $a \approx (m/m_0)^{1/2} a_0$ and 
$\rho_{\rm int} \approx (m/m_0)^{-1/2} \rho_0$ for $d_f \approx 2$ aggregates,
the corresponding radius and internal density are found to be 
\beq
a_{\rm roll} \sim 1~{\rm cm}~\pfrac{m_{\rm roll}}{10^{-4}~{\rm g}}^{1/2},
\label{eq:aroll}
\eeq
\beq
\rho_{\rm int, roll} \sim 10^{-5}~{\rm g~cm^{-3}}
~\pfrac{m_{\rm roll}}{10^{-4}~{\rm g}}^{-1/2}.
\label{eq:rhoroll}
\eeq
The triangles in Figure~\ref{fig:rho} mark the rolling mass 
at $r = 5~\AU$ and 20~AU predicted by Equation~\eqref{eq:mroll}.
The analytic prediction well explains when the decrease in $\rho_{\rm int}$ terminates.

The density evolution is more complicated at $m>m_{\rm roll}$, 
where collisional compression is no longer negligible (i.e., $E_{\rm imp} > E_{\rm roll}$).
At $r = 5~\AU$, the internal density is approximately constant until the stopping time reaches 
$\Omega t_s = 1$, and then decreases as $\rho_{\rm int} \propto m^{-1/5}$.
At $r = 20~\AU$, by contrast, the density is kept nearly constant until $m \sim 10^2~{\rm g}$
($a \sim 10^2~{\rm cm}$), and then decreases as $\rho_{\rm int} \propto m^{-1/8}$.

As shown below, the density histories mentioned above can be directly derived 
from the porosity change recipe we adopted.  
Let us assume again that aggregates grow mainly through collisions with similar-sized ones
 ($m_1\approx m_2$ and $V_1 \approx V_2$).
In this case, the evolution of $\rho_{\rm int}$ at $E_{\rm imp} \gg E_{\rm roll}$ is approximately 
given by Equation~\eqref{eq:V12_SWT08}.
Furthermore, we neglect the term $(2V_1^{5/6})^{-4}$ in Equation~\eqref{eq:V12_SWT08}
assuming that the impact energy is sufficiently large
(which is true as long as $\Omega t_s <1$; see below).
Under these assumptions, the internal density of aggregates after collision, 
$\rho_{\rm int} = 2m_1/V_{1+2}$, is approximately given by
\beq
\rho_{\rm int} \approx \pfrac{3}{5}^{3/2}
\pfrac{E_{\rm imp}}{N_{1+2}bE_{\rm roll}}^{3/10}N_{1+2}^{-1/5}\rho_0,
\label{eq:rhoint_Eimp}
\eeq
where $N_{1+2} = 2m_1/m_0$.
Since the impact energy $E_{\rm imp} \approx m_1(\Delta v)^2/4$ 
is proportional to $N_{1+2}(\Delta v)^2$, 
Equation~\eqref{eq:rhoint_Eimp} implies that 
\beq
\rho_{\rm int} \propto (\Delta v)^{3/5} m^{-1/5},
\label{eq:rhoint_compress}
\eeq
where we have dropped the subscript for mass for clarity.
Equation~\eqref{eq:rhoint_compress} gives the relation between $\rho_{\rm int}$ and $m$ 
 if we know how the impact velocity depends on them.
Explicitly, if $\Delta v \propto m^\beta \rho_{\rm int}^\gamma$, 
Equation~\eqref{eq:rhoint_compress} leads to
\beq
\rho_{\rm int} \propto m^{(3\beta-1)/(5-3\gamma)}. 
\label{eq:rhoint_m}
\eeq

In our simulation, the main source of the relative velocity is turbulence.
The turbulence-driven velocity depends on $t_s$ as $\Delta v_t \propto t_s$ at $t_s \ll t_\eta$ 
and $\Delta v_t \propto \sqrt{t_s}$ at $t_\eta \ll t_s \ll t_L(=\Omega^{-1})$
(see Equation~\eqref{eq:vT}).
As found from Equation~\eqref{eq:ts}, the stopping time depends on $\rho_{\rm int}$ and $m$ as 
$t_s \propto m/A \propto m/a^2 \propto m^{1/3}\rho_{\rm int}^{2/3}$ in the Epstein regime 
($a\ll \lambda_{\rm mfp}$) and as $t_s \propto ma/A \propto m^{2/3}\rho_{\rm int}^{1/3}$ 
in the Stokes regime ($a\gg \lambda_{\rm mfp}$). 
Using these relations with Equation~\eqref{eq:rhoint_m}, 
we find four regimes for density evolution,
\beq
\rho_{\rm int} \propto \left\{ \begin{array}{ll}
m^{0}, & a\ll \lambda_{\rm mfp}~{\rm and}~t_s \ll t_\eta, \\
m^{1/4}, & a\gg \lambda_{\rm mfp}~{\rm and}~t_s \ll t_\eta, \\ 
m^{-1/8}, & a\ll \lambda_{\rm mfp}~{\rm and}~t_\eta \ll t_s \ll t_L, \\
m^{0}, & a\gg \lambda_{\rm mfp}~{\rm and}~t_\eta \ll t_s \ll t_L. \\
\end{array}
\right.
\label{eq:rhoevol}
\eeq
 
The circles, diamonds, and square in Figure~\ref{fig:rho} mark the size at which $a = \lambda_{\rm mfp}$
 (i.e., $t_s^{\rm(Ep)} \sim t_s^{\rm(St)}$), $t_s = t_\eta$, and $\Omega t_s = 1$, respectively.
At $r = 5~\AU$, the sizes at which $a = \lambda_{\rm mfp}$ and $t_s = t_\eta$ nearly overlap, 
and hence only two velocity regimes $t_s=t_s^{\rm(Ep)} \ll t_\eta$ and 
$t_\eta \ll t_s=t_s^{\rm(St)} \ll t_L$ are effectively relevant. 
For both cases, Equation~\eqref{eq:rhoevol} predicts flat density evolution.
At $r = 20~\AU$, there is a stage in which $t_s \gg t_\eta$ and
 $a \ll \lambda_{\rm mfp}$, 
for which Equation~\eqref{eq:rhoevol} predicts $\rho_{\rm int} \propto m^{-1/8}$.
These predictions are in agreement with what we see in Figure~\ref{fig:rho}.

Equation~\eqref{eq:rhoint_Eimp} does not apply to the density evolution at $\Omega t_s > 1$, 
where the collision velocity no more increases and hence 
collisional compression becomes less and less efficient as the aggregates grow.
However, if we go back to Equation~\eqref{eq:V12_SWT08}  and assume that 
the impact energy $E_{\rm imp}$ is sufficiently small, we obtain 
$V_{1+2} \approx 2^{6/5}V_1$, or equivalently
 $V_{1+2}/m_{1+2}^{6/5} \approx V_1/m_1^{6/5}$,
where $m_{1+2} = 2m_1$ is the aggregate mass after a collision.
This implies that $V/m^{6/5}$ is kept constant during the growth, i.e., 
$V \propto  m^{6/5}$, and hence we have $\rho_{\rm int} = m/V \propto m^{-1/5}$.
This is consistent with the density evolution at $\Omega t_s > 1$ 
seen in the upper panel of Figure~\ref{fig:rho}.
  
\section{Condition for Breaking Through the Radial Drift Barrier}\label{sec:cond}
In this section, we explain why porous aggregates 
overcome the radial drift barrier in the inner region of the disk.
We do this by comparing the timescale of aggregate growth and radial drift.
We assume that dust aggregates grow mainly through collisions with similar-sized aggregates. 
As shown in Section~\ref{sec:massratio}, this is a good approximation for the growth
of aggregates dominating the total mass of the system 
(i.e., aggregates with mass $m \sim \bracket{m}_m$).
The growth rate of the aggregate mass $m$ at the midplane is then given by
\beq
\frac{dm}{dt} = \rho_{d} \sigma_{\rm coll} \Delta v
= \frac{\Sigma_d}{\sqrt{2\pi}h_d} A\Delta v,
\label{eq:dmdt}
\eeq
where $\rho_d = \Sigma_d/(\sqrt{2\pi}h_d)$ is the spatial dust density at the midplane, 
and we have approximated $\sigma_{\rm coll}$ as the projected area $A$.
Equation~\eqref{eq:dmdt} can be rewritten in terms of the growth timescale as 
\beq
t_{\rm grow}  \equiv \pfrac{d\ln m}{dt}^{-1}
= \sqrt{2\pi}\frac{h_d}{\Delta v}\frac{m/A}{\Sigma_d}
= \frac{4\sqrt{2\pi}}{3}\frac{h_d}{\Delta v}\frac{\rho_{\rm int}a}{\Sigma_d},
\label{eq:tgrow}
\eeq
where we have used that $m = (4\pi/3)\rho_{\rm int}a^3$ and $A = \pi a^2$.
What we do here is to compare $t_{\rm grow}$ with the timescale for the radial drift given by 
\beq
t_{\rm drift} \equiv  \left|\frac{d\ln r}{dt}\right|^{-1} = \frac{r}{|v_r|}.
\label{eq:tdrift}
\eeq

Now we focus on the stage at which the radial drift velocity reaches the maximum value,
i.e., $\Omega t_s =1$.
At this stage, the dust scale height is given by
 $h_d \approx (2\alpha_D/3)^{1/2}h_g$ according to Equation~\eqref{eq:hd}.
In addition, we set $\Delta v \approx \sqrt{\alpha_D}c_s$ since 
the collision velocity between $\Omega t_s = 1$ particles is dominated by the turbulence-driven velocity.
Using these relations and $h_g = c_s/\Omega$, we can rewrite Equation~\eqref{eq:tgrow} as
\beqn
t_{\rm grow}|_{\Omega t_s=1} 
&=& \frac{4}{3}\sqrt{\frac{4\pi}{3}}\frac{(\rho_{\rm int}a)_{\Omega t_s=1}}{\Sigma_d\Omega} 
\nonumber \\
&\approx& 40 \pfrac{\Sigma_d/\Sigma_g}{0.01}^{-1}
\frac{(\rho_{\rm int}a)_{\Omega t_s=1}}{\Sigma_g} t_K,
\label{eq:tgrow1} 
\eeqn
where $t_K = 2\pi/\Omega$ is the Keplerian orbital period.
Thus, the growth timescale is shorter when the mass-to-area ratio 
$m/A \propto \rho_{\rm int}a$ is smaller.
Note that $t_{\rm grow}|_{\Omega t_s=1}$ is independent of $\alpha_D$ 
since both $h_d$ and $\Delta v$ scale with $\sqrt{\alpha_D}$. 
By contrast, the drift timescale for ${\Omega t_s=1}$ particles is 
\beq
t_{\rm drift}|_{\Omega t_s=1} = \frac{1}{\eta\Omega}
\approx 40 \pfrac{\eta}{4\times10^{-3}}^{-1}t_K. 
\label{eq:tdrift1} 
\eeq
The ratio of the two timescales is
\beqn
\pfrac{t_{\rm grow}}{t_{\rm drift}}_{\Omega  t_s = 1}  &=&
\frac{4}{3}\sqrt{\frac{4\pi}{3}}\eta
\frac{(\rho_{\rm int}a)_{\Omega t_s=1}}{\Sigma_d} \nonumber \\
&\approx& 1\pfrac{\eta}{4\times10^{-3}}\pfrac{\Sigma_d/\Sigma_g}{0.01}^{-1}
\frac{(\rho_{\rm int}a)_{\Omega  t_s = 1} }{\Sigma_g}. \qquad
\eeqn

The ratio $(t_{\rm grow}/t_{\rm drift})_{\Omega  t_s = 1}$ determines 
the fate of dust growth at $\Omega t_s = 1$. 
If $(t_{\rm grow}/t_{\rm drift})_{\Omega  t_s = 1}$ is very small, dust particles grow
beyond $\Omega t_s = 1$ without experiencing significant radial drift;
otherwise, dust particles drift inward before they grow.
We expect growth without significant drift to occur if
\beq
\pfrac{t_{\rm grow}}{t_{\rm drift}}_{\Omega  t_s = 1} \la \frac{1}{30},
\label{eq:cond}
\eeq
where the threshold value $1/30$ takes into account the fact that 
$t_{\rm grow}$ is the timescale for mass doubling 
while the particles experience the fastest radial drift over decades in mass.
Below, we examine in what condition this requirement is satisfied. 
  
The ratio $(\rho_{\rm int}a)_{\Omega t_s=1}/\Sigma_g$ 
depends on the drag regime at $\Omega t_s=1$.
We consider the Epstein regime first.
Using $\rho_g = \Sigma_g/(\sqrt{2\pi}h_g)$ and $h_g =c_s/\Omega$, 
one can rewrite Epstein's law as   
$\Omega  t_s = (\pi/2) \rho_{\rm int}a/\Sigma_g$.
Thus, for $\Omega t_s =1$, we have a surprisingly simple relation 
\beq
\frac{(\rho_{\rm int}a)_{\Omega t_s=1}}{\Sigma_g} = \frac{2}{\pi}.
\label{eq:mar_Ep}
\eeq 
Inserting this relation into Equation~\eqref{eq:tgrow1},
we obtain
\beq
t_{\rm grow}|_{\Omega  t_s = 1} 
\approx 30 \pfrac{\Sigma_d/\Sigma_g}{0.01}^{-1} t_K.
\label{eq:tgrow1_Ep}
\eeq
Hence, the growth condition (Equation~\eqref{eq:cond}) 
is not satisfied for the standard disk parameters 
$\eta \approx 10^{-3}$ and $\Sigma_d/\Sigma_g = 0.01$,
in agreement with the results of our and previous \citep{BDH08} simulations.
Note that the right-hand side of Equation~\eqref{eq:tgrow1_Ep} 
is independent of $\rho_{\rm int}$.
Thus, the porosity of aggregates has no effect on the radial drift barrier within 
the Epstein regime.

The situation differs in the Stokes drag regime.
A similar calculation as above leads to  
\beq
\frac{(\rho_{\rm int}a)_{\Omega t_s=1}}{\Sigma_g} 
= \frac{9}{2\pi}\frac{\lambda_{\rm mfp}}{a|_{\Omega t_s=1}}
\label{eq:mar_St}
\eeq
and 
\beq
t_{\rm grow}|_{\Omega t_s=1} \approx 
60\pfrac{\Sigma_d/\Sigma_g}{0.01}^{-1}
\frac{\lambda_{\rm mfp}}{a|_{\Omega t_s=1}}.
\label{eq:tgrow1_St}
\eeq
Note that the growth timescale is inversely proportional to the aggregate radius, 
in contrast to that in the Epstein regime (Equation~\eqref{eq:tgrow1_Ep}) 
being independent of aggregate properties.
Substituting Equations~\eqref{eq:tdrift1} and \eqref{eq:tgrow1_St} into 
the growth condition (Equation~\eqref{eq:cond}), we find that 
aggregates break through the radial drift barrier 
in the ``deep'' Stokes regime, $a|_{\Omega t_s=1}/\lambda_{\rm mfp} \ga 45$.
Unlike Equation~\eqref{eq:tgrow1_Ep}, 
Equation~\eqref{eq:tgrow1_St} implicitly depends on $\rho_{\rm int}$
through $a_{\Omega t_s=1}/\lambda_{\rm mfp}$ (see below), 
so the porosity of aggregates does affect the growth timescale in this case.
It is interesting to note that the speedup of dust growth occurs 
even though the maximum collision velocity is the same. 
Indeed, the collision velocity depends only on $\Omega t_s$ 
and is thus independent of the drag regime. 
We remark that Stokes' law breaks down when $a$ becomes so large 
that the particle Reynolds number becomes much larger than unity,
as already mentioned in Section~\ref{sec:evol}. 
We will show in Section~\ref{sec:Newton} that this fact sets 
the minimum value of $t_{\rm grow}|_{\Omega t_s=1}$ to 
$\approx 0.3 t_K$; see Equation~\eqref{eq:tgrow1_Ne}.

\begin{figure}
\epsscale{1.1}
\plotone{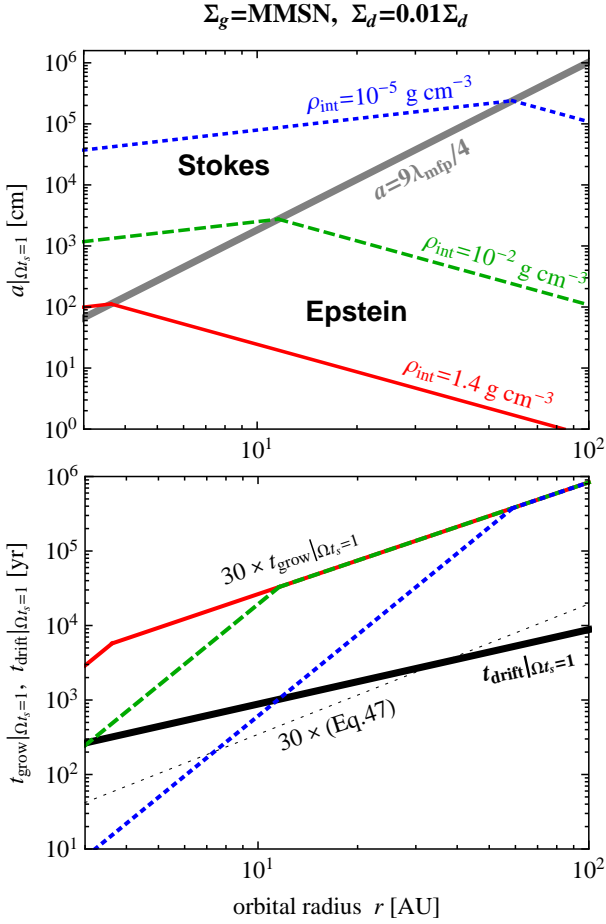}
\caption{
Size $a$ (upper panel) and growth timescale $t_{\rm grow}$ (lower panel)
of dust aggregates at $\Omega t_s = 1$ as a function of orbital radius $r$ 
for internal densities $\rho_{\rm int} = 1.4~{\rm g~cm^{-3}}$ (solid line), 
$10^{-2}~{\rm g~cm^{-3}}$ (dashed line), and $10^{-5}~{\rm g~cm^{-3}}$ (dotted line).
The MMSN with the dimensionless diffusion 
coefficient $\alpha_D = 10^{-3}$ is assumed for the disk model.
The thick line in the upper panel indicates $a = 9\lambda_{\rm mfp}/4$, 
at which the drag law changes from the Epstein regime to the Stokes regime.
The thick line in the lower panel shows the drift timescale $t_{\rm drift}$ at $\Omega t_s = 1$ (independent of $\rho_{\rm int}$).
For $\rho_{\rm int} = 10^{-5}~{\rm g~cm^{-3}}$, 
$t_{\rm grow}|_{\Omega t_s = 1}$ satisfies the growth criterion (Equation~\eqref{eq:cond}) at $r\la 10~\AU$.
In reality, $t_{\rm grow}|_{\Omega t_s = 1}$ does not fall below 
the value given by Equation~\eqref{eq:tgrow1_Ne} (thin dotted line) because of the effect 
of the gas drag at high particle Reynolds numbers (see Section~\ref{sec:Newton}).
However, this does not change the location where the growth condition is satisfied.
}
\label{fig:t}
\end{figure}
The internal density of aggregates controls the growth timescale 
through the aggregate size $a$ at $\Omega t_s = 1$.
For given $\rho_{\rm int}$, one can analytically calculate 
$a|_{\Omega t_s = 1}$ from Equations~\eqref{eq:mar_Ep} 
and \eqref{eq:mar_St}. Explicitly,
\beq
a|_{\Omega t_s = 1} = \frac{2\Sigma_g}{\pi\rho_{\rm int}}
\eeq
for the Epstein regime, and
\beq
a|_{\Omega t_s = 1} = 
\frac{3}{(2\pi)^{1/4}}\sqrt{\frac{m_gh_g}{\rho_{\rm int}\sigma_{\rm mol}}}
\label{eq:a_St}
\eeq
for the Stokes regime, where we have used 
$\lambda_{\rm mfp} = m_g/(\rho_g\sigma_{\rm mol})$
and $\rho_g = \Sigma_g/(\sqrt{2\pi}h_g)$.
For fixed $\rho_{\rm int}$, $a|_{\Omega t_s = 1}$ 
decreases with increasing $r$ in the Epstein regime, 
but increases in the Stokes regime.
The upper panel of Figure~\ref{fig:t} plots $a|_{\Omega t_s = 1}$ 
for three different values of the aggregate internal density $\rho_{\rm int}$.
If dust particles grew into compact spheres ($\rho_{\rm int} \sim 1~{\rm g~cm^{-3}}$), 
Epstein's law governs the motion of $\Omega t_s = 1$ particles 
in almost entire parts of the snow region ($r>3\AU$).
However, if dust particles grow into highly porous aggregates 
with $\rho_{\rm int} \sim 10^{-5}~{\rm g~cm^{-3}}$, 
the particles growing at $r\la 60~\AU$ enter the Stokes regime before 
$\Omega t_s$ reaches unity. 
The lower panel of Figure~\ref{fig:t} shows the two timescales $t_{\rm grow}|_{\Omega t_s=1}$
and $t_{\rm drift}|_{\Omega t_s=1}$ as calculated from Equations~\eqref{eq:tgrow1} 
and \eqref{eq:tdrift1}, respectively.
We see that compact particles with $\rho_{\rm int} \sim 1~{\rm g~cm^{-3}}$ do
not satisfy the growth condition (Equation~\eqref{eq:cond})
outside the snow line, while porous aggregates with $\rho_{\rm int} \sim 10^{-5}~{\rm g~cm^{-3}}$
do in the region $r \la 10~\AU$.
These explain our simulation results presented in Section~\ref{sec:results}.

\begin{figure}
\epsscale{1.1}
\plotone{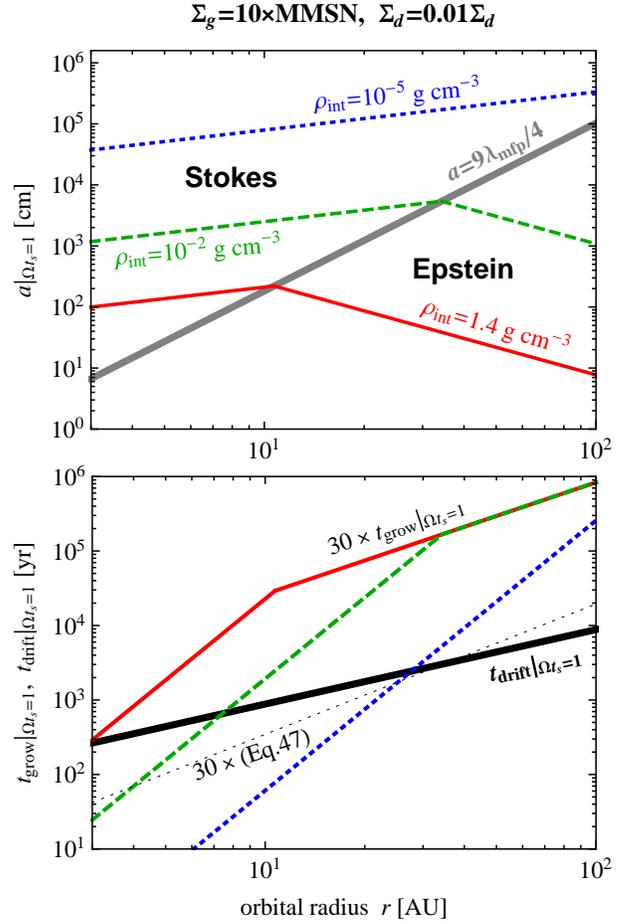}
\caption{
Same as Figure~\ref{fig:t}, but for a disk 10 times heavier than the MMSN. 
The growth criterion (Equation~\eqref{eq:cond}) is satisfied 
at $r\la 25~\AU$ for $\rho_{\rm int} = 10^{-5}~{\rm g~cm^{-3}}$ and 
at $r\la 7~\AU$ for $\rho_{\rm int} = 10^{-2}~{\rm g~cm^{-3}}$.
}
\label{fig:t_10MMSN}
\end{figure}
Finally, we remark that a high disk mass 
(i.e., a high $\Sigma_g$ with fixed $\Sigma_d/\Sigma_g$) 
favors the breakthrough of the radial drift barrier.
Figure~\ref{fig:t_10MMSN} shows the size $a$
and the timescales $t_{\rm grow}$ and 
$t_{\rm drift}$ at $\Omega t_s=1$ for a disk 10 times heavier than the MMSN.
We see that the growth condition (Equation~\eqref{eq:cond}) is now satisfied
at $r \la 25~\AU$ for $\rho_{\rm int} = 10^{-5}~{\rm g~cm^{-3}}$ and
at  $r \la 7~\AU$ even for $\rho_{\rm int} = 10^{-2}~{\rm g~cm^{-3}}$.
This is because a higher $\Sigma_g$ leads to a shorter $\lambda_{\rm mfp}$ 
and hence allows aggregates to reach the Stokes regime 
$a/\lambda_{\rm mfp} \gg 1$ at larger $r$ or with higher $\rho_{\rm int}$
(note that enhancement of $\Sigma_g$ by a constant remains 
$\eta$ and hence $t_{\rm drift}|_{\Omega t_s=1}$ unchanged). 
Interestingly, our porosity model predicts that 
$\rho_{\rm int}|_{\Omega t_s =1}$ is independent of $\Sigma_g$.
In fact, substituting Equation~\eqref{eq:a_St} with 
$(\Delta v)_{\Omega t_s=1} \approx \sqrt{\alpha_D}c_s$
and $N_{1+2} \propto \rho_{\rm int} a^3$
into Equation~\eqref{eq:rhoint_Eimp}, we obtain the equation for
$\rho_{\rm int}|_{\Omega t_s =1}$ that does not involve $\Sigma_g$.

\section{Discussion}\label{sec:discussion}
So far we have shown that the evolution of dust into highly porous aggregates is 
a key to overcome the radial drift barrier.
On the other hand, in order to clarify the role of porosity evolution, 
we have ignored many other effects relevant to dust growth in protoplanetary disks.
In this section, we discuss how the ignored effects would affect dust evolution.

\subsection{Effect of the Friction Law at High Particle Reynolds Numbers}\label{sec:Newton}
In this study, we have assumed that the stopping time $t_s$ obeys Stokes' law 
whenever $a \ga \lambda_{\rm mfp}$.
In reality, Stokes' law applies only when the particle Reynolds number 
(the Reynolds number of the gas flow around the particle)
${\rm Re}_p \equiv 2a|{\bm v}_d - {\bm v}_g|/\nu_{\rm mol}$ is less than unity,
where $|{\bm v}_d - {\bm v}_g|$ is the gas--dust relative velocity.
When ${\rm Re}_p \ga 1$, i.e., the particle becomes so large and/or the gas--dust relative velocity 
becomes so high, the stopping time becomes dependent on the particle velocity
\citep[see, e.g.,][]{W77}. 
In this subsection, we discuss how this effect affects our conclusion.

In general, the stopping time at $a \gg \lambda_{\rm mfp}$ can be written as 
\beq
t_s = \frac{2m}{C_D\rho_g|{\bm v}_d - {\bm v}_g|A},
\label{eq:ts_Ne}
\eeq
where $C_D$ is a dimensionless coefficient that depends on ${\rm Re}_p$.
Stokes' law, which applies when ${\rm Re}_p \ll 1$, is given by $C_D = 24/{\rm Re}_p$.
In the opposite limit, ${\rm Re}_p \gg 1$, the drag coefficient $C_D$ approaches a constant value
(typically of order unity; e.g., $C_D \approx 0.5$ for a sphere with  $10^3 \la {\rm Re}_p \la 10^5$),
which is known as Newton's friction law.
Thus, in the Newton regime, the stopping time depends on the particle velocity 
unlike in the Stokes regime.
In this case, one has to calculate the stopping time and particle velocity simultaneously
since the particle velocity in turn depends on the stopping time.

In the previous sections, we have ignored the Newton regime to avoid 
the above-mentioned complexity.
However, it is easy to calculate 
the growth timescale in the Newton regime for given $\Omega t_s$,
for which the gas--dust relative velocity can be known in advance.
Below, we show that the Newton drag sets 
the {\it minimum} value of  $t_{\rm grow}|_{\Omega  t_s = 1}$ (Equation~\eqref{eq:tgrow1})
for given orbital radius and internal density, 
which was not taken into account in Section~\ref{sec:cond}.
At the midplane, Equation~\eqref{eq:ts_Ne} can be rewritten as 
$\Omega t_s = (2\sqrt{2\pi}/C_D)(c_s/|{\bm v}_d - {\bm v}_g|)m/(\Sigma_g A)$,
where we have used that $\rho_g = \Sigma_g\Omega/(\sqrt{2\pi}c_s)$.
When $\Omega t_s = 1$, the gas--dust relative velocity is dominated 
by the dust radial velocity $v_r = -\eta v_K$, 
so we can set $|{\bm v}_d - {\bm v}_g| \approx \eta v_K$.
Thus, at the midplane, we obtain a relation 
\beq
\frac{(\rho_{\rm int}a)_{\Omega t_s=1}}{\Sigma_g} \approx \frac{3C_D}{8\sqrt{2\pi}}\frac{\eta v_K}{c_s}
\approx 0.07 \pfrac{C_D}{0.5}\frac{\eta v_K}{c_s},
\label{eq:mar_Ne}
\eeq
where we have used that $m/A = 4\rho_{\rm int}a/3$.
If $C_D$ reaches a constant, ${(\rho_{\rm int}a)_{\Omega t_s=1}}/{\Sigma_g}$ no longer depends on aggregate properties.
Putting this equation into Equation~\eqref{eq:tgrow1},
we have
\beq
t_{\rm grow}|_{\Omega  t_s = 1} 
\approx 0.3 \pfrac{\Sigma_d/\Sigma_g}{0.01}^{-1}\pfrac{C_D}{0.5}\pfrac{\eta v_K/c_s}{0.08} t_K.
\label{eq:tgrow1_Ne}
\eeq
When $C_D = 24/{\rm Re}_p$, Equation~\eqref{eq:tgrow1_Ne} reduces to the equation for 
the Stokes drag (Equation~\eqref{eq:tgrow1_St}), where $t_{\rm grow}|_{\Omega t_s=1}$ 
decreases with increasing aggregate size $a$.
However, when ${\rm Re}_p$ becomes so large that $C_D$ reaches a constant value,
$t_{\rm grow}|_{\Omega t_s=1}$ no longer decreases with increasing $a$.
Thus, we find that the Newton drag sets the minimum value of $t_{\rm grow}|_{\Omega t_s=1}$.
For our disk model, in which $\Sigma_d/\Sigma_g = 0.01$ and
 $\eta v_K/c_s = 0.08(r/5~\AU)^{1/4}$, 
the minimum growth timescale is $\approx 0.2$--$0.3(C_D/0.5)t_K$ at $r \approx 3$--$10~\AU$.

Since the Newton drag regime was ignored in our model, 
the growth rate of aggregates was overestimated there at high ${\rm Re}_p$.  
As seen in the lower panel of Figure~\ref{fig:t}, 
the growth timescale $t_{\rm grow}|_{\Omega t_s=1}$ 
for the $\rho_{\rm int} = 10^{-5}~{\rm g~cm^{-3}}$ aggregates 
falls below the minimum possible value given by Equation~\eqref{eq:tgrow1_Ne}
at $r\la 7~\AU$.
This implies that dust growth is somewhat artificially accelerated 
in our simulation presented in Section~\ref{sec:porous}.
However,  this artifact is {\it not} the reason why porous aggregates 
grow across the radial drift barrier in the simulation.
Indeed, the drift timescale $t_{\rm grow}|_{\Omega t_s=1}$ is $\approx 40 t_K$ 
at these orbital radii, and hence the minimum growth timescale still satisfies the condition 
for breaking through the drift barrier, Equation~\eqref{eq:cond}
(see Section~\ref{sec:cond}).
Thus, highly porous aggregates are still able to break through the radial drift barrier 
even if Newton's law at high particle Reynolds numbers is taken into account.

In summary, we have shown that Newton's friction law ($C_D \approx {\rm constant}$) at 
high particle Reynolds numbers sets a floor value for the grow timescale 
at $\Omega t_s = 1$. 
In the numerical simulation presented in Section~\ref{sec:porous},
the neglect of the Newton drag regime causes artificial acceleration of the growth 
of  $\Omega t_s \ga 1$ aggregates.
However, comparison with the drift timescale shows that the floor value of $t_{\rm grow}|_{\Omega t_s=1}$ is sufficiently small for dust to grow across $\Omega t_s = 1$.
Therefore, the deviation from Stokes' law at high particle Reynolds numbers 
has little effect on the successful breakthrough of the radial drift barrier observed
in our simulation.
\subsection{Effects of Frictional Backreaction}\label{sec:backreaction}
So far we have neglected the frictional backreaction from dust to gas
when determining the velocities of dust aggregates (Equations~\eqref{eq:vr} and \eqref{eq:vphi}). 
Here, we discuss the validity of this assumption.

\subsubsection{Effect on the Equilibrium Drift Velocity}\label{sec:NSH}
Frictional backreaction generally modifies the equilibrium velocities of both gas and dust.  
The equilibrium velocities in the presence of the backreaction are derived by 
\citet{THI05} for arbitrary dust size distribution.
The result shows that the 
radial and azimuthal velocities $v_{r}$ and $v_{\phi} = v'_{\phi} + v_K$ of dust particles 
with stopping time $t_s$ are given by 
\beq
v_r = \frac{1}{1+(\Omega t_s)^2}v_{g,r} + \frac{2\Omega t_s}{1+(\Omega t_s)^2}v'_{g,\phi}, 
\label{eq:vr_multi}
\eeq
\beq
v'_\phi =  -\frac{\Omega t_s}{2[1+(\Omega t_s)^2]}v_{g,r} + \frac{1}{1+(\Omega t_s)^2}v'_{g,\phi}, 
\label{eq:vphi_multi}
\eeq
where 
\beq
v_{g,r} = \frac{2Y}{(1+X)^2+Y^2}\eta v_K, 
\label{eq:vgasr}
\eeq
\beq
v'_{g,\phi} = -\frac{1+X}{(1+X)^2+Y^2}\eta v_K,
\label{eq:vgasphi}
\eeq
are the radial and azimuthal components of the gas velocity 
relative to the local circular Keplerian motion, respectively, and 
\beq
X = \int \frac{\rho_d(m)}{\rho_g}\frac{1}{1+(\Omega t_s(m))^2} dm,
\label{eq:X}
\eeq
\beq
Y = \int \frac{\rho_d(m)}{\rho_g}\frac{\Omega t_s(m)}{1+(\Omega t_s(m))^2} dm,
\label{eq:Y}
\eeq
with $\rho_d(m)$ being the spatial mass density of dust particles 
per unit aggregate mass.\footnote{Equations~\eqref{eq:vr_multi}--\eqref{eq:Y} are 
equivalent to the ``multi-species NSH solution'' of \citet[][their Equations~(A4) and (A5)]{BS10a}.}
In the limit of $X$, $Y \to 0$, the gas velocities approach  $v_{g,r} \to 0$ and 
$v'_{g,\phi} \to -\eta v_K$, and 
hence Equations~\eqref{eq:vr_multi} and \eqref{eq:vphi_multi}
reduce to Equations~\eqref{eq:vr} and \eqref{eq:vphi}, respectively.
Thus, the dimensionless quantities $X$ and $Y$ measure the significance of the frictional backreaction.
As found from the integrands in Equations~\eqref{eq:vr} and \eqref{eq:vphi}, 
the backreaction is nonnegligible when the local dust-to-gas mass ratio exceeds unity 
and the aggregates dominating the dust mass tightly couple to the gas.

\begin{figure}
\epsscale{1.1}
\plotone{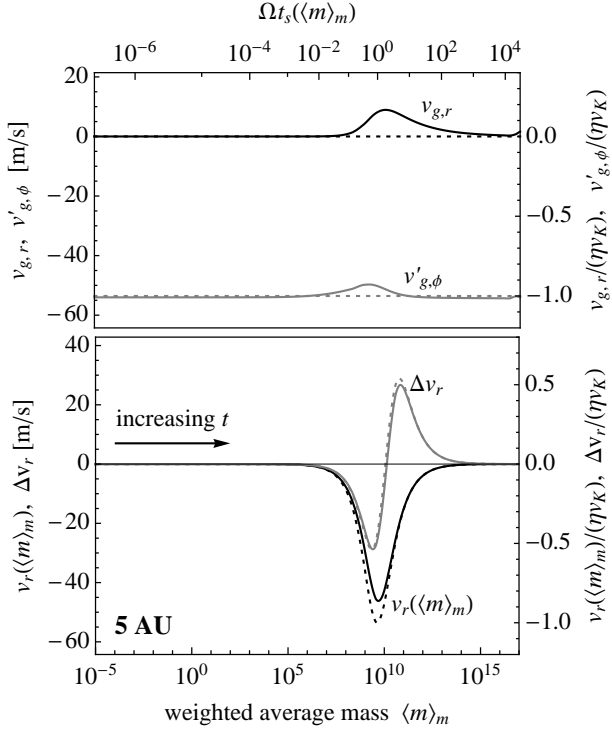}
\caption{Radial and azimuthal velocities of gas (upper panel) and radial velocity of dust (lower panel)
at $r = 5~\AU$ as a function of the weighted averaged mass $\bracket{m}_m$.
The solid black and gray curves in the upper panel show $v_{g,r}$ and $v'_{g,\phi} = v_{g,\phi}-v_K$,
respectively, obtained from the simulation including porosity evolution of aggregates 
and fractional backreaction from dust to gas.
The dotted curves are the velocities when the fractional backreaction is neglected.
}
\label{fig:vgas}
\end{figure}
To test the effect of fractional backreaction, 
we have also simulated porous aggregation 
using Equations~\eqref{eq:vr_multi} and \eqref{eq:vphi_multi}
instead of Equations~\eqref{eq:vr} and \eqref{eq:vphi} for the aggregate velocities.
However, it is found that the effect of backreaction is so small that  
the resulting dust evolution is hardly distinguishable from that presented in Section~\ref{sec:results}.
The upper panel of Figure~\ref{fig:vgas} shows the temporal evolution of the 
gas velocities $v_{g,r}$ and $v'_{g,\phi}$ observed in this simulation as a function 
of the weighted average mass $\bracket{m}_m$.
We see that the observed gas velocities deviate at most only by $9~{\rm m~s^{-1}} \approx 0.17\eta v_K$  from the velocities 
when the backreaction is absent (dotted lines).
As a result of this, the inward velocity $-v_r$ of aggregates with $m = \bracket{m}_m$ 
is decreased only by $15~\%$ even when $\Omega t_s(\bracket{m}_m) \approx 1$
(see the black solid curve in the lower panel of Figure~\ref{fig:vgas}).
The above result can be understood in the following way.
As found from the definitions of $X$ and $Y$ (Equations~\eqref{eq:X} and \eqref{eq:Y}),
the effect of the backreaction is significant only when the density of dust coupled 
to the gas ($\Omega t_s \la 1$) is comparable to or higher than the gas density.
When $\Omega t_s(\bracket{m}_m) \la 1$, the density of the coupled dust at the midplane 
is $\la \Sigma_d/h_d|_{\Omega t_s=1} \sim \Sigma_d/(h_g\sqrt{\alpha}_D) \sim 
(0.01/\sqrt{\alpha_D})\rho_{g,{\rm mid}} \sim 0.3\rho_{g,{\rm mid}} \la \rho_{g,{\rm mid}}$, 
where $\rho_{g,{\rm mid}}$ is the midplane gas density and we have used that
$h_d|_{\Omega t_s=1} \sim \sqrt{\alpha_D}h_g \sim 0.03h_g$ (Equation~\eqref{eq:hd}) 
and $\Sigma_d/\Sigma_g \approx 0.01$ (the latter is true as long as 
$\Omega t_s(\bracket{m}_m) \la 1$). 
When $\Omega t_s(\bracket{m}_m) \ga 1$, the dust density does exceed the gas density at
the midplane, but the most part of the dust mass is now carried by {\it decoupled} 
($\Omega t_s > 1$) aggregates,
which do not affect the gas motion.\footnote{Indeed, $X$ and $Y$ are insensitive 
to  $\Omega t_s \gg 1$ particles because  
the factors $1/[1+(\Omega t_s)^2] \approx  t_s^{-2}$ and 
$\Omega t_s/[1+(\Omega t_s)^2] \approx t_s^{-1}$ decrease faster
than the spatial dust density $\rho_{d} \propto \Sigma_d/h_d \propto t_s^{1/2}$ increases
(see Equation~\eqref{eq:hd})
.}
Thus, the density of coupled dust is always lower than the gas density, 
and hence the backreaction effect is insignificant at all times.

Furthermore, the effect on the {\it differential} drift velocity is even less significant,
because the decreases in the inward velocities nearly cancel out.
As an example, the gray solid and dotted curves in the lower panel of Figure~\ref{fig:vgas}
show the differential radial velocity $\Delta v_r$ between aggregates of stopping times
 $t_s = t_s(\bracket{m}_m)$ and $0.3t_s(\bracket{m}_m)$ obtained from the simulations with and without 
the backreaction, respectively.  
We see that the maximum values of $|\Delta v_r|$, 
which are reached when $\Omega t_s(\bracket{m}_m) \approx 0.7$, differ only by $5\%$.
Therefore, the frictional backreaction from dust to gas hardly affects the drift-induced collision 
velocity between dust aggregates.

\subsubsection{Streaming Instability}\label{sec:SI}
The backreaction of dust on gas causes another phenomenon, 
the so-called streaming instability \citep{YG05}.
This means that the equilibrium gas--dust motion as described 
by Equations~\eqref{eq:vr}--\eqref{eq:vphi} is unstable against perturbation.
One important consequence of this instability is rapid clumping 
of marginally decoupled ($\Omega t_s \sim 1$) dust particles \citep[e.g.,][]{JY07,J+07,BS10a}.
The clumping proceeds in a runaway manner 
(i.e., turbulent diffusion no longer limits the clumping)
once the dust density exceeds the gas density at the midplane 
(e.g., \citealt*{JY07}; see also the analytic explanation of this by \citealt*{JYM09}).  
The runaway clumps could be eventually gravitationally bound and form 100 km sized 
planetesimals \citep{J+07}.
For more tightly coupled ($\Omega t_s \ll 1$) particles, however,
the clumping occurs only moderately unless
the dust-to-gas surface density ratio is high and/or the radial drift speed is low \citep{JYM09,BS10c}. 
This is also true for loosely coupled particles ($\Omega t_s \gg 1$) for which
the interaction with the gas is weak.

As seen in Section~\ref{sec:porous}, porous aggregates 
are able to reach $\Omega t_s \sim 1$ in inner regions of disks. 
These aggregates likely trigger the streaming instability 
and can even experience runaway collapse.
However, it is not obvious whether the clumps really experience the runaway collapse, 
since the growth timescale of the $\Omega t_s \sim 1$ aggregates can be as short as 
one orbital period (see Section~\ref{sec:Newton}), which is comparable to the growth time 
of the streaming instability at $\Omega t_s = 1$ \citep{YG05}.
If the aggregates cross $\Omega t_s \sim 1$ faster than the clumps develop, 
planetesimal formation will occur via direct collisional growth 
rather than gravitational instability.
In order to address this issue, we will need to simulate 
coagulation and streaming instability simultaneously.

\subsection{Fragmentation Barrier}\label{sec:frag}
In this study, we have assumed that all aggregate collisions lead to sticking.
This assumption breaks down if the collisional velocity is so high that  
the collision involves fragmentation and erosion. 
If the mass loss due to fragmentation and erosion is significant, 
it acts as an obstacle to planetesimal formation (the so-called fragmentation barrier; e.g., \citealt*{BDH08}). 
Here, we discuss the validity and possible limitations of this assumption.

Recent $N$-body simulations predict that very fluffy aggregates made of $0.1~\micron$ sized icy particles experience catastrophic disruption
 at collision velocities $\Delta v \ga 35~{\rm m~s^{-1}}$ \citep{W+09}.
If a large aggregate grows mainly through collisions with similar-sized ones (which is true in our simulations; see Figure~\ref{fig:rate}), 
the collision velocity at $\Omega t_s \approx 1$  is dominated by the turbulence-driven velocity
$\Delta v_t \approx \delta v_g \approx \sqrt{\alpha_D}c_s$ (Section~\ref{sec:dv}).
If the disk is optically thin and moderately turbulent ($\alpha_D = 10^{-3}$) as in our model,
the collision velocity is $\approx 21~{\rm m~s^{-1}}$ at $r=5~\AU$,
so catastrophic disruption is likely insignificant for such collisions.
However, if turbulence is as strong as $\alpha_D = 10^{-2}$, 
the collision velocity at $r=5~\AU$ and $\Omega t_s = 1$ goes up to $67~{\rm m~s^{-1}}$.
In protoplanetary disks, strong turbulence with $\alpha_D \ga 10^{-2}$ can be driven by 
magnetorotational instability (MRI; e.g., \citealt*{BH98}).
If such strong turbulence exists, fragmentation becomes no more negligible even for icy aggregates. 
Besides, the collision velocity can become higher than the above estimate when a large aggregate collides 
with much smaller ones, 
since the collision velocity is then dominated by the radial drift motion. 
For example, the differential radial drift velocity between an $\Omega t_s = 1$ aggregate and 
a much smaller one is as high as $\approx \eta v_K \approx 56~{\rm m~s^{-1}}$ in
optically thin disks.
At such a high velocity, erosion by small aggregates can also slow down the growth of 
$\Omega t_s \approx 1$ aggregates, although net growth might be possible 
\citep[see, e.g.,][]{TW09,TKW11}.

On the other hand, resupply of small dust particles by fragmentation/erosion has
positive effects on dust growth. 
First, small dust particles stabilize MRI-driven turbulence because 
they efficiently capture ionized gas particles
and thereby reduce the electric conductivity of the gas \citep[e.g.,][]{SMUN00}.
This process generally leads to the reduction of the gas random velocity (and hence the reduction 
of turbulence-induced collision velocity),
especially when the magnetic fields threading the disk are weak \citep{OH11}.
In addition, small fragments enhance the optical thickness of the disk, and thus reduce the temperature of the gas in the interior of the disk 
(given that turbulence is stabilized there).
Since the radial drift velocity is proportional to the gas temperature, this leads to the reduction of the 
drift-induced collision velocity.
In the limit of large optical depths, the gas temperature is reduced by a factor 
$\approx (h/r)^{1/4} \approx 0.5$ near the midplane \citep{KNH70}, resulting in the reduction of the 
drift-induced collision velocity to $28~{\rm m~s^{-1}}$.
These effects may help the growth of large aggregates beyond the fragmentation barrier.
 
The size of monomers is another key factor.
Although we have assumed monodisperse monomers of $a_0 = 0.1\micron$, 
the size of interstellar dust particles ranges from nanometers to microns.
It is suggested both theoretically \citep{CTH93,DT97} and experimentally \citep{BW08} 
that the threshold velocity for sticking is roughly inversely proportional to $a_0$.
Thus, inclusion of larger monomers generally leads to the decrease in
the sticking efficiency.
However, it is not obvious whether aggregates composed of multi-sized interstellar particles are mechanically weaker or stronger than aggregates considered in this study. 
For example, if the monomer size distribution $dn_0/da_0$ obeys that 
of interstellar dust particles, $dn_0/d\log a_0 \propto a_0^{-5/2}$ \citep{MRN77}, 
the total mass of the aggregates is dominated by the largest ones
($m_0\propto a_0^3$ and hence $m_0dn_0/d\log a_0 \propto a_0^{1/2}$). 
Nevertheless, the existence of smaller monomers can still be  important, 
since the binding energy per contact 
$E_{\rm break}$ is proportional to $a_0^{4/3}$ \citep{CTH93,DT97}
and hence the total binding energy tends to be dominated by the smallest ones
($E_{\rm break}dn_0/d\log a_0 \propto a_0^{-7/6}$).
The net effect of  multi-sized monomers needs to be clarified  
by future numerical as well as laboratory experiments.

Another issue about the growth efficiency of icy aggregates arises from sintering. 
Sintering is redistribution of ice molecules on solid surfaces due to vapor transport and other effects. 
In this process, ice molecules tend to fill dipped surfaces (i.e., surfaces with negative surface curvature) since the equilibrium vapor pressure decreases with decreasing the surface curvature. 
In an aggregate composed of equal-sized icy monomers, this process leads to 
growth of the monomer contact areas \citep{S11b} and consequently to 
enhancement of the aggregate's mechanical strength such as $F_{\rm roll}$.
Significant growth of the contact areas could cause the reduction of 
the aggregate's sticking efficiency
since the dissipation of the collision energy through internal 
rolling/sliding motion could then be suppressed \citep{S99}.   
Furthermore, if the monomers have different sizes, sintering 
leads to evaporation of smaller monomers (having higher positive curvature), 
which may result in the breakup of the aggregate \citep{S11a}.
Therefore, sintering can prevent the growth of icy aggregates
near the snow line where sintering proceeds rapidly.
\citet{S11b} shows that the timescale of ${\rm H_2O}$ sintering falls below $10^3$ yr 
in the region between the snow line ($3~\AU$) and $7~\AU$ for the radial temperature 
adopted in our study.   
This is comparable to the timescale on which submicron-sized icy particles 
grow into macroscopic objects in this region (see Figure~\ref{fig:5AU}).
However, if the disk is passive and optically thick \citep{KNH70}, 
no icy materials (including ${\rm H_2O}$ and ${\rm CO_2}$) 
undergo rapid sintering at $r \ga 4~\AU$ \citep{S11b}. 
Moreover, the required high optical depth can be provided 
by tiny fragments that would result from the sintering-induced fragmentation itself.
Consistent treatment of the two competing effects 
is necessary to precisely know the location where sintering is really problematic.

To summarize, whether icy aggregates survive catastrophic fragmentation and erosion crucially depends on
the environment of protoplanetary disks as well as on the size distribution of the aggregates and constituent monomers.
However, we emphasize that icy aggregates can survive within a realistic range of disk conditions as explained above.
Indeed, the range is much wider than that for rocky aggregates, for which catastrophic disruption occurs 
at collision velocities as low as a few ${\rm m~s^{-1}}$ \citep{BW08,W+09,G+10}.
In order to precisely predict in what conditions icy aggregates overcome the fragmentation barrier,
we need to take into account the mass loss due to fragmentation/erosion and the reduction of collision velocities 
due to the resupply of small particles in a self-consistent way.
This will be done in our future work.

\subsection{Validity and Limitations of the Porosity Model}\label{sec:porosity_validity}
Aggregates observed in our simulation have very low internal densities.
This is a direct consequence of the porosity model we adopted (Equation~\eqref{eq:V12}).
Here, we discuss the validity and limitations of our porosity model.
 
As mentioned in Section~\ref{sec:porosity}, our porosity change recipe at 
$E_{\rm imp} \ga E_{\rm roll}$ is based on head-on collision experiments 
of similar-sized aggregates.
In our simulation, dust growth is indeed dominated by collision with similar-sized aggregates
(see Section~\ref{sec:massratio}), 
so our result is unlikely affected by the limitation of the porosity model regarding 
the size ratio.
By contrast, the neglect of offset collision may cause underestimation of the porosity increase,
since the impact energy is spent for stretching rather than compaction at offset collision \citep{W+07,PD09}.
If this is the case, then the breakthrough of the radial drift barrier can occur even outside 
$10~\AU$.

On the other hand, the formation of low-density dust aggregates 
is apparently inconsistent with the existence of massive and much less porous aggregates 
in our solar system.
For example, comets, presumably the most primitive dust ``aggregates'' in the solar system,
are expected to have 
mean internal densities of $\rho_{\rm int} \sim 0.1~{\rm g~cm^{-3}}$ \citep[e.g.,][]{GH90}.
Since our porosity model does not explain the formation of such large and less porous ``aggregates,'' there should exist any missing compaction mechanisms.  

One possibility is {\it static} compression due to gas drag and self-gravity.
Although static compression is ignored in our porosity model, 
it can contribute to compaction of aggregates that are massive or 
decoupled from the gas motion.
For relatively compact ($\rho_{\rm int} \sim 0.1~{\rm g~cm^{-3}}$) 
dust cakes made of micron-sized ${\rm SiO_2}$ particles,
static compaction is observed to occur at static pressure $>100~{\rm Pa}$ 
\citep{BS04,G+09}.
By contrast, the static compression strength has not yet been
measured so far for icy aggregates with very low internal densities 
($\rho_{\rm int} \ll 0.1~{\rm g~cm^{-3}}$).
However, for future reference, it will be useful to estimate here the static pressures 
due to gas drag and self-gravity.

The ram pressure, the gas drag force per unit area, 
is given by $P_{\rm ram} = C_D\rho_g|{\bm v}_d-{\bm v}_g|^2/2$,
where $C_D$ is the drag coefficient and $|{\bm v}_d-{\bm v}_g|$ is the gas--dust relative speed
(see Section~\ref{sec:Newton}).
At $\Omega t_s \ga 1$, the gas--dust relative speed is approximately equal to $\eta v_K$.
Thus, assuming Newton's drag law $C_D \sim 1$ for $\Omega t_s \ga 1$ aggregates
 (Section~\ref{sec:Newton}), the ram pressure at $\Omega t_s \ga 1$ is estimated as
\beq
P_{\rm ram} \sim \rho_g(\eta v_K)^2 
\sim 10^{-5}\pfrac{\rho_g}{10^{-11}~{\rm g~cm^{-3}}}\pfrac{\eta v_K}{50~{\rm m~s^{-1}}}^2 ~{\rm Pa}
\label{eq:P1}
\eeq
independently of aggregate properties.
Thus, if the static compression strength of our high porous aggregates is 
lower than $10^{-5}~{\rm Pa}$, 
compression of the aggregates will occur at $\Omega t_s \ga 1$ due to ram pressure.   

The static pressure due to self-gravity is estimated from dimensional analysis 
as
\beq
P_{\rm grav} \sim \frac{Gm^2}{a^4} 
\sim 10^{-7} \pfrac{m}{10^{10}~{\rm g}}^{2/3}
\pfrac{\rho_{\rm int}}{10^{-5}~{\rm g~cm^{-3}}}^{4/3} ~{\rm Pa}.
\eeq
For $m \sim 10^{10}~{\rm g}$ and $\rho_{\rm int} \sim 10^{-5}~{\rm g~cm^{-3}}$, 
which correspond to the $\Omega t_s = 1$ aggregates 
observed in our simulation (Figure~\ref{fig:evolP}),
the gravitational pressure is much weaker than the ram pressure.
However, since $P_{\rm grav} \propto m^{2/3}$, 
compression due to self-gravity becomes important for much heavier aggregates.
For example, if $\rho_{\rm int}\sim  10^{-5}(m/10^{10}~{\rm g})^{-1/5}~{\rm g~cm^{-3}}$
as is for the $\Omega t_s \ga 1$ aggregates observed in our simulation,
$P_{\rm grav}$ exceeds $P_{\rm ram}$ at $m \sim 10^{17}~{\rm g}$, 
which is comparable to the mass of comet Halley.
Moreover, since $P_{\rm grav} \propto \rho_{\rm int}^{4/3}$, 
gravitational compaction will proceed in a runaway manner unless
 the static compression strength increases more rapidly than $P_{\rm grav}$.
Thus, static compression due to self-gravity may be a key to fill the gap between 
our high porous aggregates and more compact planetesimal-mass bodies in the solar system.

\section{Summary and Outlook}\label{sec:summary}
We have investigated how the porosity evolution of dust aggregates affects 
their collisional growth and radial inward drift.
We have applied a porosity model based on $N$-body simulations of aggregate collisions \citep{SWT08,SWT12}.
This porosity model  allows us to study the porosity change upon collision for a wide range of impact energies.
As a first step, we have neglected the mass loss due to collisional fragmentation 
and instead focused on dust evolution outside the snow line, 
where aggregates are mainly composed of ice and hence catastrophic 
fragmentation may be insignificant \citep{W+09}. 
Our findings are summarized as follows.

\begin{enumerate}
\item
Icy aggregates can become highly porous even if collisional compression 
is taken into account (Section~\ref{sec:porous}).
Our model calculation suggests that the internal density of icy aggregates 
at $5~\AU$ falls off to $10^{-5}~{\rm g~cm^{-3}}$ by the end of the initial fractal growth stage
and then is kept to this level until the aggregates decouple from the gas motion (Figure~\ref{fig:rho}).
Stretching of merged aggregates at offset collisions, which is not taken into account in our porosity model, 
could further decrease the internal density \citep{W+07,PD09}.

\item
A high porosity triggers significant acceleration in collisional growth.
This acceleration is a natural consequence of particles' 
aerodynamical property in the Stokes regime, i.e., at particle radii 
larger than the mean free path of the gas molecules (Section~\ref{sec:cond}).
The porosity (or internal density) of an aggregate determines whether the aggregate 
reaches the Stokes regime before the radial drift stalls its growth.
Compact aggregates tend to drift inward before experiencing the rapid growth, 
while highly porous aggregates are able to experience it over a wide range of 
the orbital radius (Figure~\ref{fig:t}).
 
\item
The growth acceleration enables the aggregates 
to overcome the radial drift barrier in inner regions of the disks.
Our model calculation shows that the breakthrough 
of the radial drift barrier can occur at orbital radii less than $10~\AU$ 
in the MMSN (Figure~\ref{fig:evolP}).
A higher disk mass allows this to occur at larger orbital radii 
or higher internal densities (Figure~\ref{fig:t_10MMSN}).
The radial drift barrier has been commonly thought to be one of the most serious obstacles 
against planetesimal formation. 
Our result suggests that, if the fragmentation of icy aggregates is truly insignificant (see Section~\ref{sec:frag}), 
formation of icy planetesimals is possible via direct collisional growth of submicron-sized icy particles
even without an enhancement of the initial dust-to-gas mass ratio.

\item
Further out in the disk, the growth of porous icy aggregates is still limited by the radial drift barrier, 
but their inward drift results in enhancement of the dust surface density 
in the inner region (Figure~\ref{fig:sigmaP}). 
This enhancement may help the core of giant planets to form within a disk lifetime \citep{KTKI10,KTK11}. 
\end{enumerate}
 
We remark that the quick growth in the Stokes regime was also observed in
recent coagulation simulations by \citet[][see their Figure~11]{BDB10} 
and \citet[][see their Figure 3]{Z+11}.
\citet{BDB10} observed the breakthrough of the radial drift barrier only at 
small orbital radii ($r\la0.5~\AU$) since they assumed compact aggregation.
\citet{Z+11} found rapid growth of porous aggregates in the Stokes regime, 
but did not consider the loss of the dust surface density through radial drift.
What we have clarified in this study is that porosity evolution indeed enables the 
breakthrough of the radial drift barrier at much lager orbital radii.
 
The porosity evolution can 
even influence the evolution of solid bodies after planetesimal formation.
It is commonly believed that the formation of protoplanets begins with the 
runaway growth of a small number of planetesimals due to gravitational focusing 
\citep[e.g.,][]{WS89}.
The runaway growth requires a sufficiently high gravitational escape 
velocity $v_{\rm esc} = \sqrt{2Gm/a}$ relative to the collision velocity. 
Since the escape velocity decreases with decreasing internal density 
($v_{\rm esc} \propto m^{1/3}\rho_{\rm int}^{1/6}$), 
it is possible that a high porosity delays the onset of the runaway and 
thereby affects its outcome.
For example, a recent protoplanet growth model including 
collisional fragmentation/erosion \citep{KTKI10,KTK11} suggests that planetesimals 
need to have grown to $>10^{21}~{\rm g}$ before the runaway growth begins
in order to enable the formation of gas giant planets
within the framework of the core accretion scenario \citep{M80,P+96}.
The size of the ``initial'' planetesimals can even determine 
the mass distribution of asteroids in the main belt \citep{M+09,W11}.
As we pointed out in Section~\ref{sec:porosity_validity}, 
compaction of large and massive aggregates may occur through 
static compression due to gas drag or self gravity. 
To precisely determine when it occurs is beyond the scope of this work, but 
it will be thus important to understand later stages of planetary system formation. 
We will address this in future work.

\acknowledgments
The authors thank the anonymous referee for useful comments.
We also thank Tilman Birnstiel, Frithjof Brauer, Cornelis Dullemond, Shu-ichiro Inutsuka,
Chris Ormel, Taku Takeuchi, Takayuki Tanigawa, Fredrik Windmark, 
and Andras Zsom for fruitful discussions.  
S.O.~acknowledges support by Grants-in-Aid for JSPS Fellows ($22\cdot 7006$) 
from MEXT of Japan.




\end{document}